\documentclass[aps, pra, showpacs, nofootinbib,10pt, longbibliography,twocolumn, amsfonts, amsmath, amssymb, superscriptaddress, floatfix]{revtex4-2}

\usepackage{hyperref}

\usepackage[T1]{fontenc}
\usepackage{color}
\usepackage{graphicx}
\usepackage{epsfig}
\usepackage[normalem]{ulem}
\usepackage{braket}
\usepackage{comment}
\def\be{\begin{equation}}
\def\ee{\end{equation}}
\def\bea{\begin{eqnarray}}
\def\eea{\end{eqnarray}}

\def\bi{\begin{itemize}}
\def\ei{\end{itemize}}

\usepackage{tikz}
\newcommand*\emptycirc[1][0.4ex]{\tikz\draw (0,0) circle (#1);} 
\newcommand*\fullcirc[1][0.4ex]{\tikz\fill (0,0) circle (#1);} 

\begin{document}

\title{Recent progress on quantum simulations of non-standard Bose-Hubbard models}

\author{Titas Chanda}
\affiliation{Department of Physics, Indian Institute of Technology Indore, Khandwa Road, Simrol, Indore 453552, India}

\author{Luca Barbiero} 
\affiliation{Institute for Condensed Matter Physics and Complex Systems,
DISAT, Politecnico di Torino, I-10129 Torino, Italy}

\author{Maciej Lewenstein}
\affiliation{ICFO-Institut de Ciencies Fotoniques, The Barcelona Institute of
Science and Technology, Castelldefels (Barcelona) 08860, Spain}  
\affiliation{ICREA, Pg. Llu\'is Companys 23, 08010 Barcelona, Spain} 

\author{Manfred J. Mark}
\affiliation{Institut f\"ur Quantenoptik und Quanteninformation, \"Osterreichische Akademie der Wissenschaften, Innsbruck, Austria}
\affiliation{Universit\"at Innsbruck, Institut f\"ur Experimentalphysik, Innsbruck, Austria}

 \author{Jakub Zakrzewski}
\affiliation{ Instytut Fizyki Teoretycznej, Wydzia\l{} Fizyki, Astronomii i Informatyki Stosowanej, Uniwersytet Jagiello\'nski, \L{}ojasiewicza 11, PL-30-348 Krak\'ow, Poland} 
\affiliation{ Mark Kac Complex Systems Research Center, Jagiellonian University in Krak\'ow, PL-30-348 Krak\'ow, Poland} 

\date{\today}
\begin{abstract}
In recent years, the systems comprising of bosonic atoms confined to optical lattices at ultra-cold temperatures have demonstrated tremendous potential to unveil novel quantum mechanical effects appearing in lattice boson models with various kinds of interactions. In this progress report, we aim to provide an exposition to recent advancements in quantum simulations of such systems, modeled by different `non-standard’ Bose-Hubbard models, focusing primarily on long-range systems with dipole-dipole or cavity-mediated interactions.
Through a carefully curated selection of topics, which includes the emergence of  quantum criticality beyond Landau paradigm, bond-order wave insulators, the role of interaction-induced tunneling, the influence of transverse confinement on observed phases, or the effect of cavity-mediated all-to-all interactions, we report both theoretical and experimental developments from the last few years. Additionally, we discuss the real-time evolution of systems with long-range interactions, where sufficiently strong interactions render the dynamics non-ergodic. And finally to cap  our discussions off, we survey recent experimental achievements in this rapidly evolving field, underscoring its interdisciplinary significance and potential for groundbreaking discoveries.
\end{abstract}

\maketitle

\tableofcontents

\section{Introduction}
Since its derivation in 1963~\cite{Hubbard63}, the Hubbard model became an iconic Hamiltonian in condensed matter physics~\cite{Lieb2003,Essler2005}. As originally conceived, this model mimics electrons in a discrete geometry characterized by different two-body interactions: pair- and density-induced-tunneling processes, as well as inter-site and onsite density-density interaction. Nevertheless, the peculiar screened shape of the Coulomb potential led to a drastic simplification of this model where the inter-particle interactions reduce to onsite processes only. After almost three decades, this last point in addition to the lack, at that time, of concrete physical implementations influenced the derivation of the bosonic counterpart of the Hubbard Hamiltonian: the Bose-Hubbard model (BHM)\cite{Fisher1989} capturing the motion of bosonic particles interacting through contact repulsion. Let us note, that the original derivation of the lattice boson model should be credited to Gersch and Knollman~\cite{Gersch63} - this work preceded that of Hubbard.

In such a context, the end of twentieth century represents the beginning of a new and exciting era where the advent of atomic quantum simulators working at ultra-low temperatures~\cite{Jaksch98} promised, among other things, to revolutionize the way that Hubbard models were studied and conceived~\cite{Bloch2008,Lewenstein12}. This promise became reality in the early 2000s, when a new technology, the optical lattices, enabled the first experimental realization of the Bose-Hubbard model~\cite{Greiner02}. This revolutionary result allowed for the blossoming of new ideas~\cite{Lahaye09,Eckardt2017} to experimentally realize non-standard BHMs~\cite{Dutta15}, characterized by beyond onsite interacting terms, which can result in the appearance of new states of matter. As shown by the first realizations of BHMs with density-induced tunneling~\cite{Jurgensen2014} and density-density inter-site interaction~\cite{Baier2016}, nowadays non-standard BHMs~\cite{Dutta15} can be realized efficiently, paving the way for a new adventure in the exploration of strongly correlated quantum matter.

Nine years have passed since the appearance of a previous review on extended Hubbard models~\cite{Dutta15}. The aim of this report in progress is to provide an update on this rapidly developing field covering some of the advancements that occurred in the last decade in the engineering and characterization of novel non-standard BHMs. The progress is stimulated by the realization that non-standard BHMs yield thrilling possibilities to demonstrate novel states of matter, often with quite intriguing properties. More importantly, however, is the enormous recent experimental progress with modifying and manipulating optical lattices, harnessing and controlling inter-atomic interactions, and, in particular, recent success with experimental implementations of long-range interactions either of dipolar nature or those mediated by resonant cavities.

The progress in this areas over the last few years is quite broad and spectacular. Instead of providing a comprehensive review of various aspects of the field, we have chosen to describe a few carefully chosen examples to illustrate the recent progress. To make this work self contained we introduce the basics of the tight binding description in Section~\ref{sec:school}. Section~\ref{den-den} discusses ground state properties coming from different non-standard BHMs discussing e.g., topological quantum criticality, the role of interaction induced tunnelings or how the transverse confinement may profoundly affect the phase diagram of non-standard BHMs. The next Section~\ref{sec:cav} reviews phases for cavity-mediated interactions, again of current experimental interest. We do not restrict ourselves, however, to ground state physics, we discuss possible nontrivial dynamics occurring at ``infinite'' temperature i.e., for initial high-energy states in Section~\ref{sec:dyn}. Importantly, we do not forget about the real excitement in this field describing the current status of leading experiments (Section~\ref{sec:exp}). We conclude by speculating on the possible future developments as well as we mention topics that we had to omit in this, necessarily, brief report.  

\section {Construction of the ``lattice'' representation}
\label{sec:school}

The usual derivation of a Bose-Hubbard Hamiltonian starts by considering a system of ultracold bosons trapped in an optical lattice having wave-number ${k_0 = 2\pi/\lambda_0}$ with $\lambda_0/2 = a$ being the lattice constant. In particular, we focus on scenarios where atoms can either move in the two-dimensional (2D) $x$-$z$ plane or in a one-dimensional (1D) line along the $z$-direction. The single-particle Hamiltonian of the atoms is given by
\begin{equation}
\mathcal{H}^0_{sp} = -\frac{\hbar^2}{2m}\nabla^2 + V_0 \left( \cos^2(k_0 z) + \beta \cos^2(k_0 x) \right) + \frac{m \omega^2}{2} y^2,
\label{eq:Hsp}
\end{equation}
where the first term corresponds to the kinetic energy of the atoms and the second term denotes the potential seen by the atoms due to the optical lattice in $x$-$z$ plane with $V_0$ and $\beta V_0$ being the lattice depths in $z$ and $x$ directions respectively. For the standard isotropic 2D lattice system we have $\beta = 1$, while $\beta \gg 1$ corresponds to the 1D case. Unless stated otherwise, we predominantly keep our discussion concentrated on 1D optical lattice systems throughout this progress report. The lattice depth $V_0$ is typically measured in the units of the lattice recoil energy $E_R = \hbar^2 k_0^2/2m$, e.g., $s=V_0/E_R$. The additional harmonic potential in the third term, with $m\omega^2/2 \gg V_0$, restricts the atomic motion in the $y$-direction. The resulting many-body Hamiltonian, in the second-quantization language, that takes into account two-body interactions, reads as~\cite{Jaksch98}
\begin{align}
\hat{H}=&\int d^3\mathbf{r} \ \hat{\Psi}^\dagger (\mathbf{r}) \ \mathcal{H}_{sp}^0 \  \hat{\Psi}(\mathbf{r})\nonumber\\
+ & \frac{1}{2}\int d^3\mathbf{r} \int d^3\mathbf{r}' \ \hat{\Psi}^\dagger(\mathbf{r})\hat{\Psi}^\dagger (\mathbf{r}')U_{\rm int}(\mathbf{r}-\mathbf{r}')\hat{\Psi}(\mathbf{r}')\hat{\Psi}(\mathbf{r}) \ ,  \label{secquant}
\end{align}
where the field operators $\hat{\Psi}(\mathbf{r})$ and $\hat{\Psi}^\dagger(\mathbf{r})$ obey the commutation relation $\left[\hat{\Psi}(\mathbf{r}), \hat{\Psi}(\mathbf{r}')^\dagger \right]=\delta ^3\left(\mathbf{r}-\mathbf{r}' \right)$, while the specific form of the two-body interaction, $U_{\rm int}(\mathbf{r}-\mathbf{r}')$, depends on the experimental setup.
 
The field $\hat{\Psi}(\mathbf{r})$/${\hat{\Psi^\dagger}(\mathbf{r})}$ operators may be decomposed in terms of annihilation/creation bosonic operators $\hat b_{\mathbf{j}}$/$\hat b^\dagger_{\mathbf{j}}$,  labeled by the lattice site $\mathbf{j} = (j_x, j_z)$ using the lowest-band Wannier function $W_{\mathbf{j}}(\mathbf{r})$ localized at position ${\mathbf{r}_{\mathbf{j}} = (a j_x, a j_z)}$ as $\hat{\Psi}(\mathbf{r}) = \sum_{\mathbf{j}} W_{\mathbf{j}}(\mathbf{r}) \hat{b}_{\mathbf{j}}$~\cite{Jaksch98}.
In a 1D geometry, after evaluating the integrals (for details see, e.g.,~\cite{Kraus20,fraxanet2022}) and keeping the most relevant terms, one obtains the general description of the system in Eq. \eqref{secquant}:
\begin{align}
\hat{H}=&-t\sum_{j=1}^{L-1}\left(\hat{b}^\dagger_j\hat{b}_{j+1}+{\rm H.c.}\right) +\frac{U}{2}\sum_{j=1}^L \hat{n}_j\left(\hat{n}_j-1\right)\nonumber\\
&+ \frac{1}{2}\sum_{j\ne k}^{L}V_{|j-k|}\hat{n}_{j}\hat{n}_{k}\nonumber\\
&-T\sum_{j=1}^{L-1}\left[\hat{b}^\dagger_j\left( \hat{n}_j+\hat{n}_{j+1}\right)\hat{b}_{j+1}+{\rm H.c.}\right]\nonumber\\
&+\frac{P}{2} \sum_{j=1}^{L-1}\left(\hat{b}^\dagger_{j+1}\hat{b}^\dagger_{j+1} \hat{b}_j\hat{b}_j +{\rm H.c.}\right).
\label{ham}
\end{align}
The first line gives the standard one-dimensional Bose-Hubbard Hamiltonian~\cite{Fisher1989} with tunneling amplitude $t$ and onsite interaction strength $U$. Note that we take into account the nearest neighbor tunneling only. This approximation is valid for sufficiently deep optical lattices, say $s>5$ (compare~\cite{Trotzky12}). 
The second line gives inter-site interactions
$V_{|j-k|}$ 
whose possible shapes will be discussed in the next sections. The third line in Eq.~\eqref{ham} denotes the interaction-induced tunnelings (IIT), while the last line describes pair tunneling processes. We point out that in all the models and setups covered in this review, these last terms result in $P\approx 0$. Nevertheless, as  described in~\cite{Li_2016}, pair tunnelings might be of fundamental relevance in order to investigate the physics of $p$-orbital models. In this regard, we underline that proposals to explore such systems in atomic quantum simulators have been very recently derived \cite{DiLiberto2023,goldman2023}.


\section{Ground state physics for the non-standard Bose-Hubbard model}
\label{den-den}

\begin{figure}
    \centering
    \includegraphics[width = 0.8\linewidth]{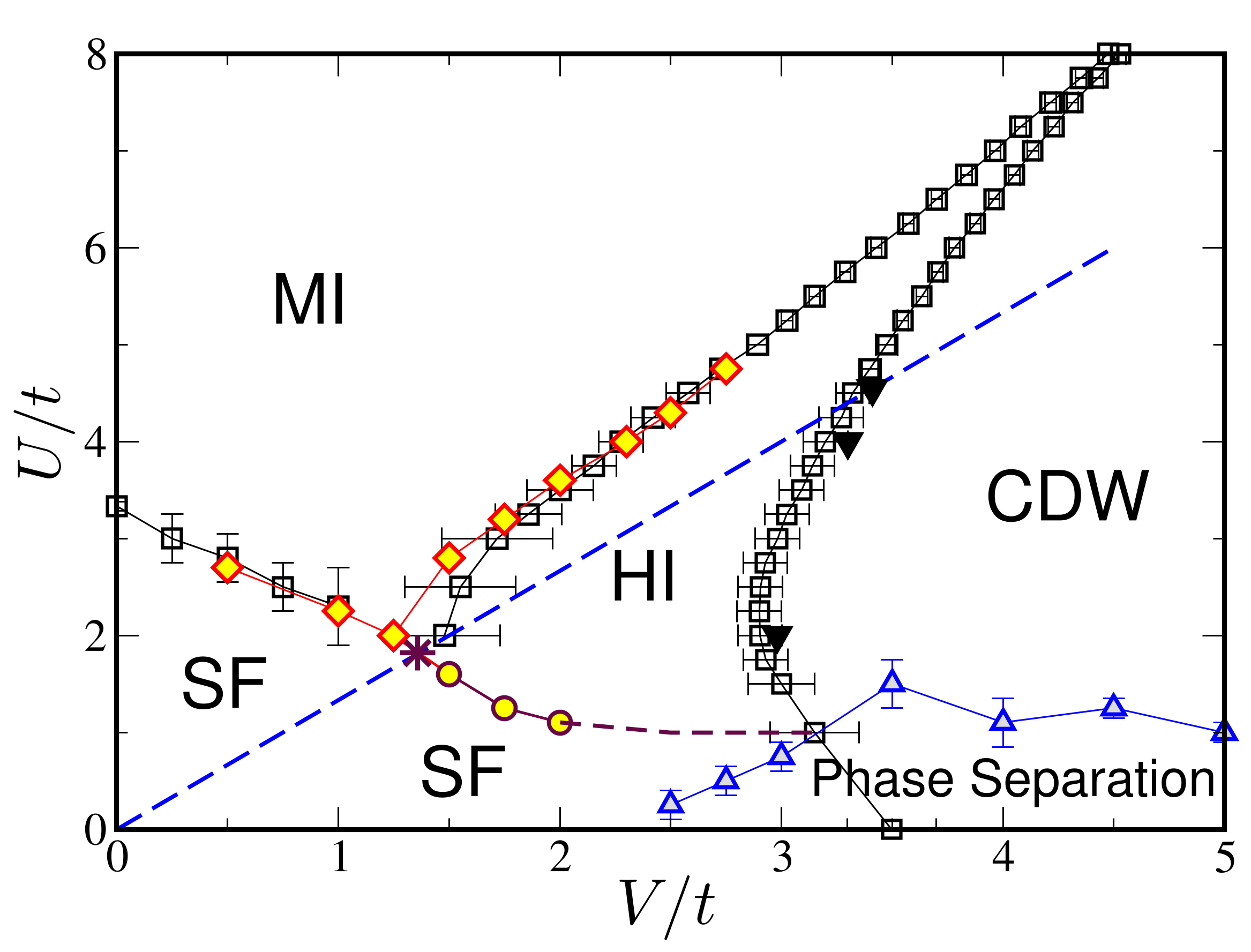}
    \caption{The phase diagram of the `standard' EBH model at unity filling in the $(U/t, V/t)$-plane obtained by quantum Monte Carlo (QMC) calculations. The figure has been adapted and reprinted with permissions from~\cite{Batrouni2014} published in 2014 by the American Physical Society.}
    \label{fig:ebh}
\end{figure}

Before considering the non-standard effects, let us first review the physics of the `standard' extended Bose-Hubbard (EBH) model, i.e., the Hamiltonian \eqref{ham} in the regime where $T=P=0$, and $V_1 = V$ while $V_{|j-k|} = 0$ for $|j-k| > 1$. While in higher dimensions such EBH models have been investigated, both theoretically~\cite{Trefzger_2011} and experimentally~\cite{Baier2016,su2023dipolar}, to capture the presence of states of matter with broken translational symmetry, i.e., charge-density wave (CDW) and supersolid (SS) phases, along with standard Mott insulator (MI) and superfluid (SF) phases, its 1D version has provided an incomparable resource to investigate symmetry protected topological (SPT)~\cite{Pollmann2010} phases. As first pointed out in~\cite{Torre06} and subsequently confirmed in a series of papers~\cite{Berg2008, Deng2011,Dalmonte2011, Rossini12,Batrouni2013,Batrouni2014, Ejima2015}, for intermediate and comparable values of the onsite interaction strength $U$ and the inter-site interaction amplitude $V$ the system supports a topological phase, called the Haldane insulator (HI), at the unit density regime $\rho=N/L=1$ (with $N$ being the total particle number and $L$ being the number of lattice sites), see Sec.~\ref{sec:crit_topo}. The phase diagram of the EBH model, at unit filling, is shown in Fig.~\ref{fig:ebh}. Apart from the different phases mentioned above, a region of phase separation (PS) appears for smaller $U$ and large $V$~\cite{Batrouni2014, Kraus22}.

Let us also mention shortly that interesting physics may occur also for fillings larger than unity. For example, filling 3/2 allows one to find low-lying excitations as fractional domain walls between different charge-density waves. This domain walls may be interpreted as non-Abelian Fibonacci anyons~\cite{Wikberg12,Duric17}. While of limited importance in 1D, when combined into a two-dimensional network, braiding of Fibonacci anyon excitations has potential applications for fault tolerant, universal, topological quantum computation. 

\subsection{Order Parameters}

To distinguish different phases appearing in the EBH models, several observables have been considered in literature. For example, to determine the spectral properties of the system one can calculate the so-called bulk neutral gap $\Delta_n$ and the charge gap $\Delta_c$ defined as~\cite{Batrouni2013, Batrouni2014}
\begin{align}
    \Delta_n &= E_1(N, L) - E_{0}(N, L), \nonumber \\
    \Delta_c &= E_{0}(N+1, L) + E_{0}(N-1, L) - 2 E_{0}(N, L),
    \label{eq:gaps}
\end{align}
where $E_{0}(N, L)$ and $E_1(N, L)$ are the ground-state and first excited-state energies respectively for a system of length $L$ with particle number $N$. Finite values of the charge gap $\Delta_c$
implies the presence of an insulting state such as MI, CDW, or HI, while on the other hand, for the gapless phases (e.g., the standard SF phase) both  gaps vanish in the thermodynamic limit~\footnote{It should be noted that for the spontaneous symmetry broken (SSB) phases, such as the CDW, the above definition of $\Delta_n$ vanishes due to ground-state degeneracy. In such cases, the neutral gap is defined after adding a symmetry breaking perturbation  to the Hamiltonian by hand so that it remains finite for gapped SSB phases.
}.

Since in 1D the gapless phases are Luttinger liquids~\cite{Berg2008, Cazalilla2011, Giamarchi2003}, the off-diagonal correlation 
\begin{equation}
    C_j(r)=\braket{\hat{b}^{\dagger}_j \hat{b}_{k}}
    \label{corr}
    \end{equation}
is expected to strictly decay algebraically with the distance $r = |j-k|$. Moreover, any modulations in the off-diagonal correlations can be revealed by the momentum distribution being defined as the averaged Fourier transform of $C_j(r)$~\cite{Jiang12}:
\begin{align}
    M(q) = \frac{1}{L^2} \sum_{j, r} e^{-i q r} C_j(r).
    \label{eq:Mq}
\end{align}
For a standard SF phase the maximum component of $M(q)$ is at $q=0$, while for other gapless phases with modulations in the off-diagonal correlations, like the staggered SF (SSF) discussed below, $M(q)$ can attain the maximum value for specific non-zero values of $q$. 

\begin{table*}[htb]
    \centering
    \begin{tabular}{|c|c||c|c|c|c|c|c|c|}
    \hline
         Phase &  Acronym & $\Delta_n$ & $\Delta_c$ & $M(q)$ & $S(q)$ & $\mathcal{O}_S^z$ & $\mathcal{O}_S^x$ \\
          &   & (Eq.~\eqref{eq:gaps}) & (Eq.~\eqref{eq:gaps}) &  (Eq.~\eqref{eq:Mq}) & (Eq.~\eqref{eq:Sq}) &  (Eq.~\eqref{eq:string}) &  (Eq.~\eqref{eq:string})\\
          \hline
         Mott insulator & MI & $\neq 0$ & $\neq 0$ & $ = 0$ & $= 0$ & $=0$ & $=0$  \\
        Superfluid & SF & $=0$ & $= 0$ & $ \neq 0 \ (q=0)$ & $= 0$ & $=0$ & $=0$  \\
        Charge-density wave & CDW & $ \neq 0$ & $\neq 0$ & $= 0$ & $\neq 0 \ (q \neq 0)$ & $\neq 0$ & $=0$  \\
        Haldane insulator & HI & $\neq 0$ & $\neq 0$ & $= 0$ & $=0 $ & $\neq 0$ & $\neq 0$  \\
        Supersolid & SS & $=0$ & $= 0$ & $ \neq 0 \ (q=0)$ & $\neq 0 \ (q \neq 0)$ & $\neq 0$ & $=0$  \\
        Staggered superfluid & SSF & $=0$ & $= 0$ & $ \neq 0 \ (q \neq 0)$ & $= 0$ & $=0$ & $=0$ \\
        staggered supersolid & SSS & $=0$ & $= 0$ & $ \neq 0 \ (q \neq 0)$ & $\neq 0 \ (q \neq 0)$ & $\neq 0$ & $=0$  \\
        \hline
    \end{tabular}
    \caption{Different phases and their acronyms appearing in `non-standard' EBH systems with corresponding values of the observables in the thermodynamic limit.}
    \label{tab:obs}
\end{table*}

Density modulated phases, i.e., regimes where the translational symmetry of the system is broken, can be distinguished by the diagonal density-density correlations~\cite{Rossini12, Berg2008} and their Fourier transform:
\begin{equation}
    S(q) = \frac{1}{L^2} \sum_{j, k} e^{i q (j-k)} \braket{\hat{n}_j \hat{n}_k},
    \label{eq:Sq}
\end{equation}
the so-called structure factor. As an example, it is straightforward to understand that for two-site translational symmetry broken phases $S(q)$ shows a peak at $q = \pi$. Such density-modulated phases, where $S(q)$ attains a maximum at $q \neq 0$, can be either a CDW or a SS phase depending on whether the phase is gapped or a superfluid.  

The topological nature of the HI phase is uniquely captured by the long-range behavior of non-local string correlation functions~\cite{Torre06, Berg2008, Batrouni2013, Batrouni2014}:
\begin{equation}
    \mathcal{O}_S^{\alpha}(r) = \braket{\hat{S}^{\alpha}_j \ e^{i \pi \sum_{k=j}^{j+r} \hat{S}^{\alpha}_k} \ \hat{S}^{\alpha}_{j+r}} \qquad \alpha=x,z,
    \label{eq:string}
\end{equation}
where $\hat{S}^x_j = \frac{1}{\sqrt{2}}\big(\sqrt{1-\frac{\hat{n}_j}{2}} \hat{b}_j + \hat{b}_j^\dagger\sqrt{1-\frac{\hat{n}_j}{2}}\big)$ and $\hat{S}^z_j= \hat{n}_j - \rho$. In the HI phase, both $\mathcal{O}_S^z$ and $\mathcal{O}_S^x$ remain finite as $r \rightarrow \infty$.

Phases and corresponding values of these observables are summarized in Table~\ref{tab:obs}.

\subsection{Topological quantum criticality}
\label{sec:crit_topo}

The observation that the string correlations $\mathcal{O}_S^{\alpha}, \alpha = x, z$ show long-range behaviors in the HI phase has made it possible, on one side, to reveal the SPT nature of this regime and, on the other, to establish a rigorous connection between this EBH Hamiltonian and the spin-$1$ Heisenberg model~\cite{Haldane1983}, whose topological nature has already been deeply understood~\cite{Pollmann2010}. It has become further clear that larger values of $V$ renders the HI phase unstable. In particular, a strong inter-site repulsion makes it possible for the breaking of the translational symmetry and therefore for the appearance of a CDW spontaneous symmetry broken (SSB) phase characterized by the perfect alternation between pairs of bosons and empty sites. Such specific symmetry breaking implies that the string correlation function $\mathcal{O}_S^z(r)$ as well as the two-point correlator ${\cal{C}}(r)=\langle S^z_j S^z_{j+r}\rangle$  shows long-range behavior, while the other string correlation function $\mathcal{O}_S^x(r)$ decays exponentially with the distance $r$.
 
 \begin{figure}
\includegraphics[width=\linewidth]{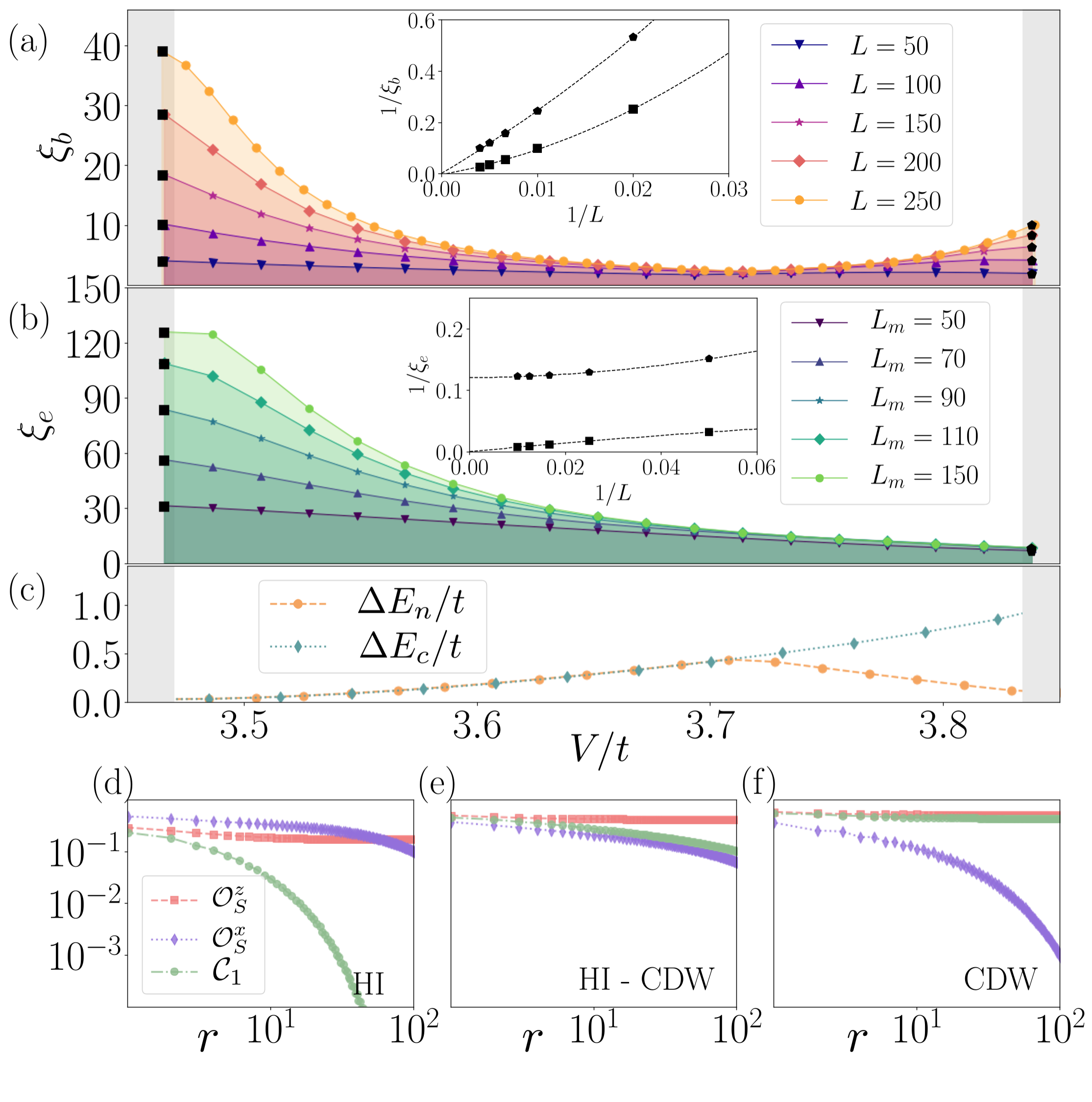}
\caption{(a) The bulk $\xi_b$ and (b) the edge $\xi_e$ correlation
lengths in units of the lattice spacing $a$ as a function of $V/t$ for the EBH model at unit filling
with
$U/t=6$. The insets show the
finite size extrapolation of $\xi_e$ and $\xi_b$ at the the MI-HI (squares) and
HI-CDW (pentagons) critical points where $L_m$ is the maximum length
to extract $\xi_e$.
(c) The neutral gap $\Delta E_{n}$ (orange) and the charge gap $\Delta E_{c}$ (blue) for
$L=200$. The gaps are computed by fixing the edge occupation by
means of large chemical potential. (d)-(f) The decay
of ${\cal{C}}(r)$ (green), ${\cal{O}}^x_S(r)$ (purple) and ${\cal{O}}^z_S(r)$ (magenta) relative to: (d) HI at
$V/t=3.65$, (e) the HI-CDW critical point at $V/t=3.86$, and (f)
CDW at $V/t=3.91$. 
The figure has been adapted and reprinted with permissions from~\cite{fraxanet2022} published in 2022 by the American Physical Society.
}
\label{tqcp}
\end{figure}
 
Recently, an important question has been put forward, wondering whether topological properties can persist at  critical points not captured by the Landau's theory \cite{Landau_ssb} like the ones involving a topological phase. 
In order to shed light on this subject, a recent matrix-product state (MPS)~\cite{Schollwck2011, Ors2014, Paeckel2019} based analysis in~\cite{fraxanet2022} has explored the behavior of both the bulk $\xi_{b}$ and the edge $\xi_e$ correlation lengths, see Fig. \ref{tqcp}(a) and (b), and of both the bulk and charge gaps as reported in Fig. \ref{tqcp}(c). Specifically, it has been shown that $\Delta E_n$, vanishes at the critical point between the HI and CDW phases and, as a consequence, the bulk correlation length $\xi_{b}\sim \Delta^{-1}_n$ has been found to diverge, see Fig. \ref{tqcp}(a).
 As, in general, SPT phases are expected to occur in presence of a finite gap, the previous results were pointing in the direction that the topological properties of the HI are lost at this critical point. Nevertheless, the same analysis has revealed that the charge gap, $\Delta E_{c}$,
presented in Fig. \ref{tqcp}(c), remains finite at this transition point. Moreover, it has been discovered that the edge states of the HI phase are still finite at the critical point as pointed out by the stable and finite value of the edge correlation length $\xi_e$\footnote{$\xi_e$ is extracted from a linear fit of $\log(|E_+-E_-)$ versus $L$, where $E_{\pm}$ are the energies of the two degenerate ground states $|L\rangle\pm|R\rangle$ with $|L\rangle(|R\rangle)$ denoting a state with the
left (right) edge state occupied by a bosonic pair and the right (left) edge state empty} depicted in Fig. \ref{tqcp}(b). Finally, the long-range character of ${\cal{O}}_S^{z}(r)$ along with the algebraic decay of ${\cal{O}}_S^{x}(r)$ and ${\cal{C}}(r)=\langle S^z_j S^z_{j+r}\rangle$ reported in Fig.~\ref{tqcp}(d)-(f) allowed one to identify this critical point as new SPT regime called topological quantum critical point (TQCP) -- previously uniquely predicted to occur in a spin-$1$ chain~\cite{Verresen2021}. Furthermore, the same investigation in~\cite{fraxanet2022} has also shown the possible appearance of TQCPs in a Hubbard chain subject to a lattice dimerization along with inter-site repulsion.


\subsection{Frustrated extended Bose-Hubbard model}
\label{sec:FEBH}

\begin{figure}
\includegraphics[width=0.8\linewidth]{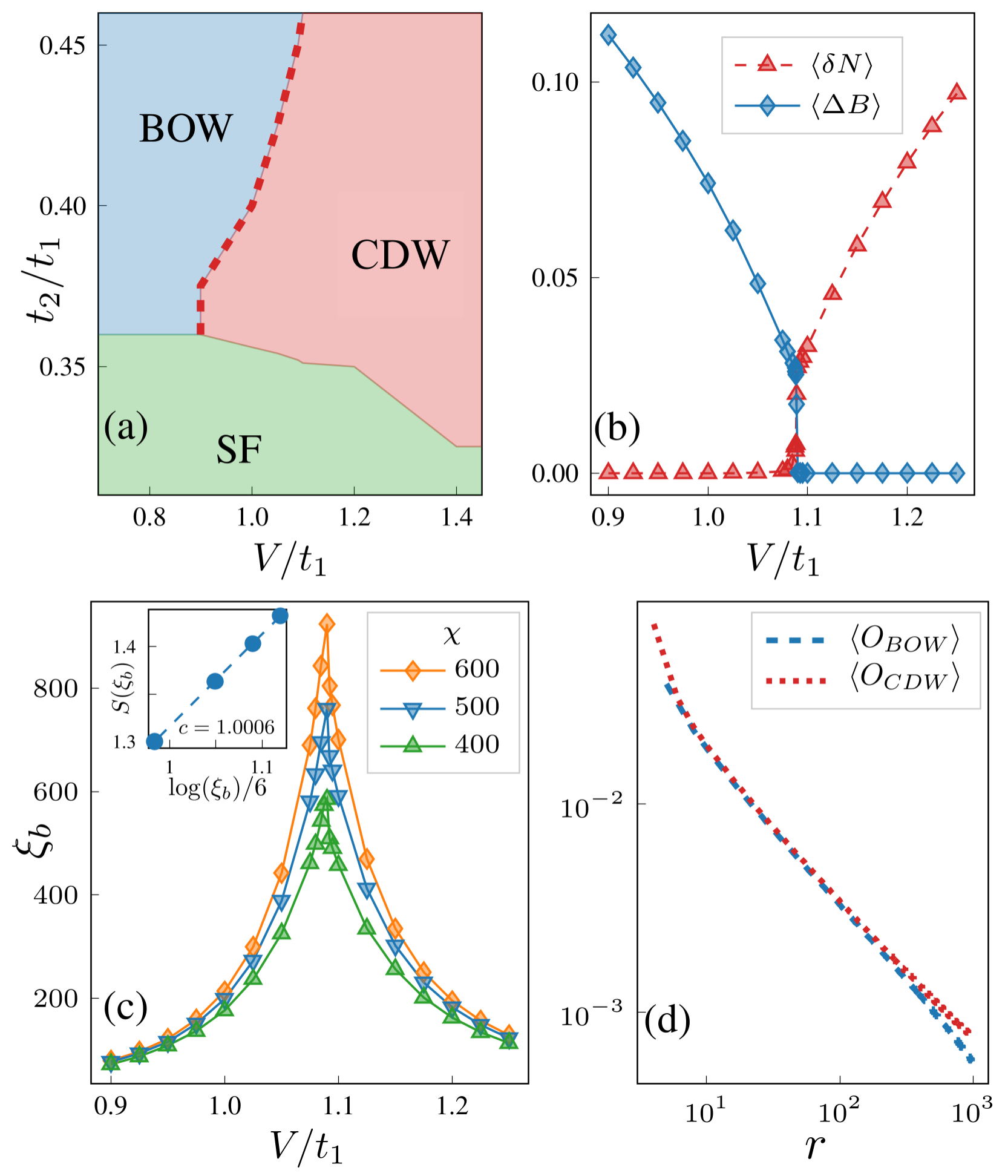}
\caption{Effect of the nearest-neighbor repulsion $V$ in the FEBH model described by the Hamiltonian~\eqref{ham2}. For all the panels, we fix $U/t_1=6$ and density $\rho = 1/2$. (a) Phase diagram of $H_{\text{FEBH}}$ in the $(V/t_1, t_2/t_1)$ plane. (b) $\Delta B$ and $\delta N$ as a function of $V/t_1$ for $t_2/t_1=0.45$. (c) The bulk correlation length $\xi_b$  as a function of $V/t_1$ for different MPS bond dimensions $\chi$ and fixed $t_2/t_1=0.45$. Inset: scaling of the entanglement entropy $S(\xi_b)$ as a function of $\xi_b$ at the critical point for bond dimensions $\chi = 400,500,600$ showing the extrapolated central charge $c=1$. (d) Decay of correlation functions $O_{\text{CDW}}(r) = \braket{(\hat{n}_j - \rho) (\hat{n}_{j+r} - \rho)}$ and $O_{\text{BOW}}(r) = \braket{(\hat{B}_{j} + \hat{B}_{j+1})(\hat{B}_{j+r} + \hat{B}_{j+r+1})}$ at the critical point for fixed $t_2/t_1=0.45$.
The figure has been adapted and reprinted with permissions from~\cite{baldelli2023} published in 2024 by the American Physical Society.
}
\label{figFQM}
\end{figure}

Effective geometrical frustration can be naturally generated in the EBH model by specifically tuning the sign of the hopping processes and by enlarging their range up to next-nearest neighbor sites. An example of a frustrated EBH (FEBH) Hamiltonian fulfilling such constraints is the one derived in~\cite{baldelli2023}:
\begin{align}
    \hat{H}_{\text{FEBH}}=&-\sum_{j}\Big[t_2(\hat{b}^{\dagger}_j \hat{b}_{j+2} + \text{H.c.}) + t_1(-1)^j(\hat{b}^{\dagger}_j \hat{b}_{j+1} + \text{H.c.})\Big] \nonumber\\
    &+\frac{U}{2}\sum_{j} \hat{n}_j(\hat{n}_j-1)+V\sum_j \hat{n}_j \hat{n}_{j+1}.
\label{ham2}
\end{align}
Here, $t_2$ refers to the the tunneling processes connecting sites spaced by two lattice sites, while the frustration is induced by the staggered sign of $t_1$. Contrary to previous proposals~\cite{Eckardt_2010,Greschner2013,Zhang2015bis,Cabedo2020,Roy2022,Halati2023,Barbiero2023} to generate frustration in Bose-Hubbard models, the implementation of Eq. \eqref{ham2} does not require neither Floquet procedures to tune the hopping sign nor direct realizations of frustrated geometries. Specifically, it can be achieved through optical lattices at the anti-magic wavelength~\cite{Karski2009,Groh2016,Anisimovas2016} where, depending on the atom polarizability, bosons can be effectively trapped both in the maxima and the minima of the optical lattice. As a result, the effective lattice spacing is reduced by a factor two, i.e., $\lambda/4$, with respect to usual optical lattices. This point allows to induce strong inter-site interactions not only through dipolar couplings but, when possible~\cite{baldelli2023}, also by tuning the scattering length to large values. The phase diagram of the model in Eq. \eqref{ham2} is reported in Fig. \ref{figFQM}(a).

Here, as expected, for low $t_2$ and $V$ a normal superfluid occurs. On the other hand, the results in Fig. \ref{figFQM} show that at the specific density $\rho=1/2$ the presence of frustration gives rise to a SSB bond-order-wave (BOW) insulating phase captured by the local order parameter\footnote{Notice that the $+$ in the definition of $\Delta B$ between the two operators is required because of the specific gauge constraint in which we are working, namely by the staggered $t_1$.}
\begin{equation}
 \Delta B = \frac{1}{L}\sum_j\langle \hat{B}_j + \hat{B}_{j+1}\rangle,   
\end{equation}
 with $\hat{B}_j=(\hat{b}_j^\dagger \hat{b}_{j+1} + \hat{b}_{j+1}^\dagger \hat{b}_{j})$ and where $\Delta B \neq 0$ implies the breaking of the discrete site inversion symmetry. By increasing $V$, the BOW is replaced by a CDW phase where, as already pointed out, the discrete translational symmetry is broken, which can be captured by the density modulation
 \begin{equation}
     \braket{\delta N} = \frac{1}{L} \sum_j (-1)^j(\braket{\hat{n}_j} - \rho).
 \end{equation}
The Landau-Wilson-Ginsburg paradigm of phase transitions~\cite{Landau_ssb,wilson_ssb} states that, as a long as two different SSB are connected through a phase transition, this has to be discontinuous where, therefore, the gap never vanishes. Nevertheless, the analysis in~\cite{baldelli2023} shown in Fig. \ref{figFQM} demonstrated instead this phase transition to be continuous. This is confirmed by the fact that the local order parameters relative to the BOW and CDW clearly vanish continuously at the same transition point, see Fig. \ref{figFQM}(b). As a consequence, at this transition point the gap is expected to vanish and therefore to support the diverging correlation length reported in Fig. \ref{figFQM}(c).  In addition, in Fig. \ref{figFQM}(d), the presence of a critical transition point is further confirmed by the algebraic decay of the correlators capturing the BOW and CDW ordering. These results allow the identification of this transition point as a deconfined quantum critical point~\cite{senthil2023deconfined}, whose existence has been first predicted in two dimensional frustrated quantum magnets~\cite{Senthil2004} and systems with multi-spin interactions~\cite{Sandvik2007}, and subsequently extended to systems in both lower~\cite{Jiang2019,Roberts2019} and higher dimensions~\cite{Charrier2008,Sreejith2015}. Their relevance lies in the fact that on one side they represent an example of transition points totally induced by quantum fluctuations and therefore not captured by the Landau-Wilson-Ginsburg paradigm and, on the other, they can be characterized by fractional excitations and emergent gauge fields.

\subsection {Interaction induced tunnelings}

Let us now enrich the problem a bit by considering the role of interaction induced tunnelings (IIT) in the Hamiltonian \eqref{ham}, i.e., we consider the situation with non-zero $T$ coefficient.  The role of IIT terms for contact interactions was discussed in detail in~\cite{Dutta15} here we shall concentrate more on physics of dipolar interactions for which IIT may strongly affect the phase diagram. 

In the case particles have strong dipolar momenta, $U_{\text{int}}(\mathbf{r})$ consists of both, the contact interaction term $U_c(\mathbf{r}) = g \delta^{(3)}(\mathbf{r})$ with $g = 4 \pi \hbar^2 a_s/m$ and $a_s$ being the $s$-wave scattering length, and the dipolar term
 \begin{align}
U_d(\mathbf{r})=\frac{C_{dd}}{4\pi} \frac{1-3\cos^2(\theta)}{r^3} \,,
\end{align}
where $\theta$ is the angle between the dipole and $\mathbf{r}$, and $C_{dd}$ is either $\mu_{0}\mu_m^{2}$ for particles having a permanent magnetic dipole moment $\mu_m$ ($\mu_{0}$ being the permeability of vacuum) or $\mu_e^{2}/\varepsilon_{0}$ for particles having a permanent electric dipole moment $\mu_e$ ($\varepsilon_{0}$ being the vacuum dielectric constant). We shall use a dimensionless quantity $d = m C_{dd}/(2\pi^3\hbar^2a)$ to characterize the dipole interaction strength.
We further assume that $V_{|j-k|} = V/a^3|j-k|^3$ in \eqref{ham}, seemingly correct for permanent dipole-dipole interactions. 

For shallow enough lattices, as mentioned above, one may consider even next-neighbor tunnelings, see e.g.,~\cite{Bartolo13,Biedron18}. However, even with nearest-neighbor hopping, the ground state of the Hamiltonian \eqref{ham} is very rich, as it may be affected by several parameters: $t$, $U$, $V_{|i-j|}$, and $T$, which depend in turn on the optical lattice geometry and depth as well as on the mutual strengths of the dipolar and contact interactions. These have to be determined with care for any experimental realization, as the physics of the model strongly depends on them. We shall just give a few examples here.

 \begin{figure}
\includegraphics[width=\linewidth]{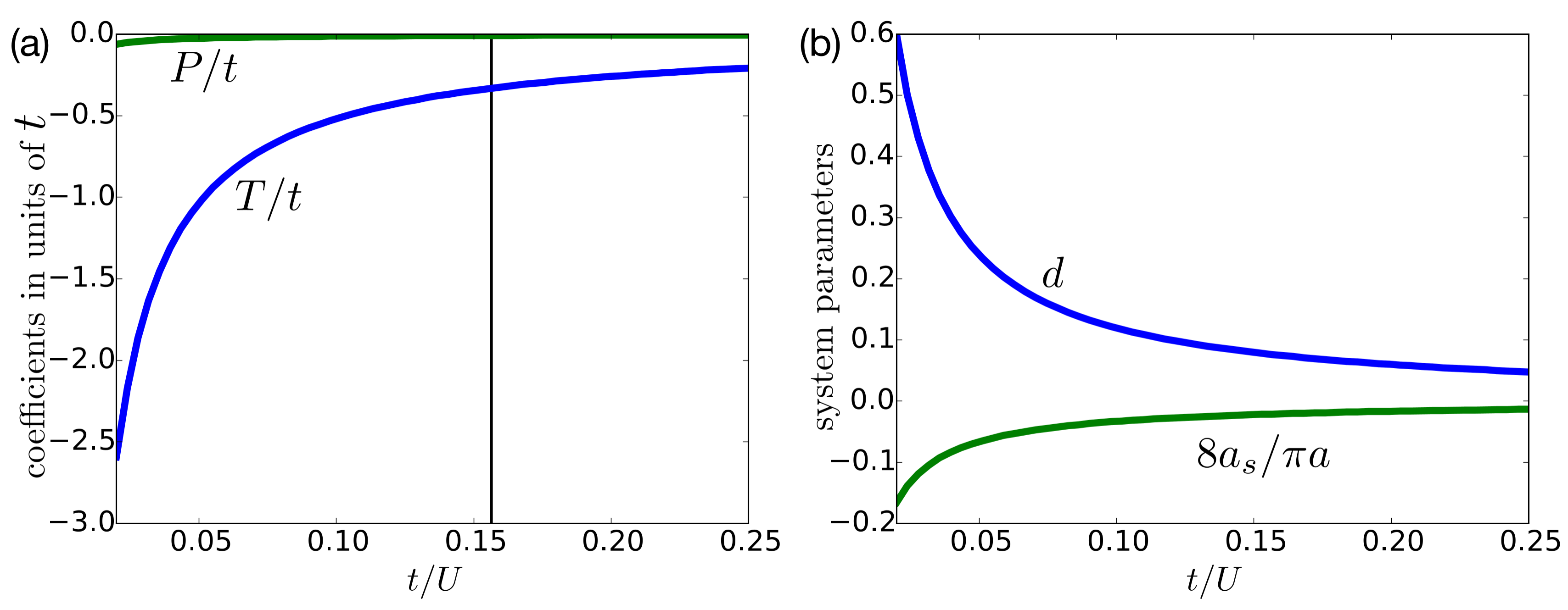}
\caption{(a) Interaction induced tunneling coefficient $T$ (blue) and pair-hopping coefficient $P$ (green) as a function of $t/U$ for $V/U=0.5$. (b)  The corresponding values of the dimensionless dipolar interaction strength $d$ (blue) and of the $s$-wave scattering length $a_s$ in units of $a$ (green) are shown. By changing both $d$ and $a_s$ the ratio $V/U$ is changed. The vertical black line in (a) indicates the value of $t/U$ for which kinetic and interaction induced tunnelings mutually cancel in the mean-field consideration for uniform density $\rho=2$.
The figure has been adapted and reprinted with permissions from~\cite{Kraus20} published in 2022 by the American Physical Society.}
\label{params}
\end{figure}

Figure~\ref{params}(a) 
shows the relations between typical parameters appearing in the Hamiltonian for 1D lattices (assumed depth $s=8$) obtained when varying the dipolar strength $d$ and the scattering length $a_s$ as shown in Fig.~\ref{params}(b). In effect $V/U=0.5$ is kept constant here. Note that the IIT coefficient $T$ is negative and is of the same order as the kinetic tunneling $t$. 
This immediately suggests a possible negative interference between both mechanisms of particle motion. Writing the tunneling terms together as~\cite{Luehmann2012}:
\begin{equation}
 \hat {\mathcal T}_{\rm eff}=\sum_{j}\hat{b}^\dagger_j\hat{b}_{j+1}\left[-t-T(\hat n_j+\hat n_{j+1}-1)\right] +{\rm H.c.}\,
 \label{eq:tuneff}
\end{equation}
one observes that for density $\rho$ the mean-field value of this term vanishes for $-t-T(2\rho-1)=0$. For $\rho=2$ this leads to the condition $3T=-t$ , visualized in Fig.~\ref{params}(a) by the dotted line. The IIT will strongly modify the phase diagram as discussed below. On the other hand the pair-tunneling coefficient $P$ is 2 orders of magnitude smaller and thus can be neglected.
 
\begin{figure}
\includegraphics[width=0.9\linewidth]{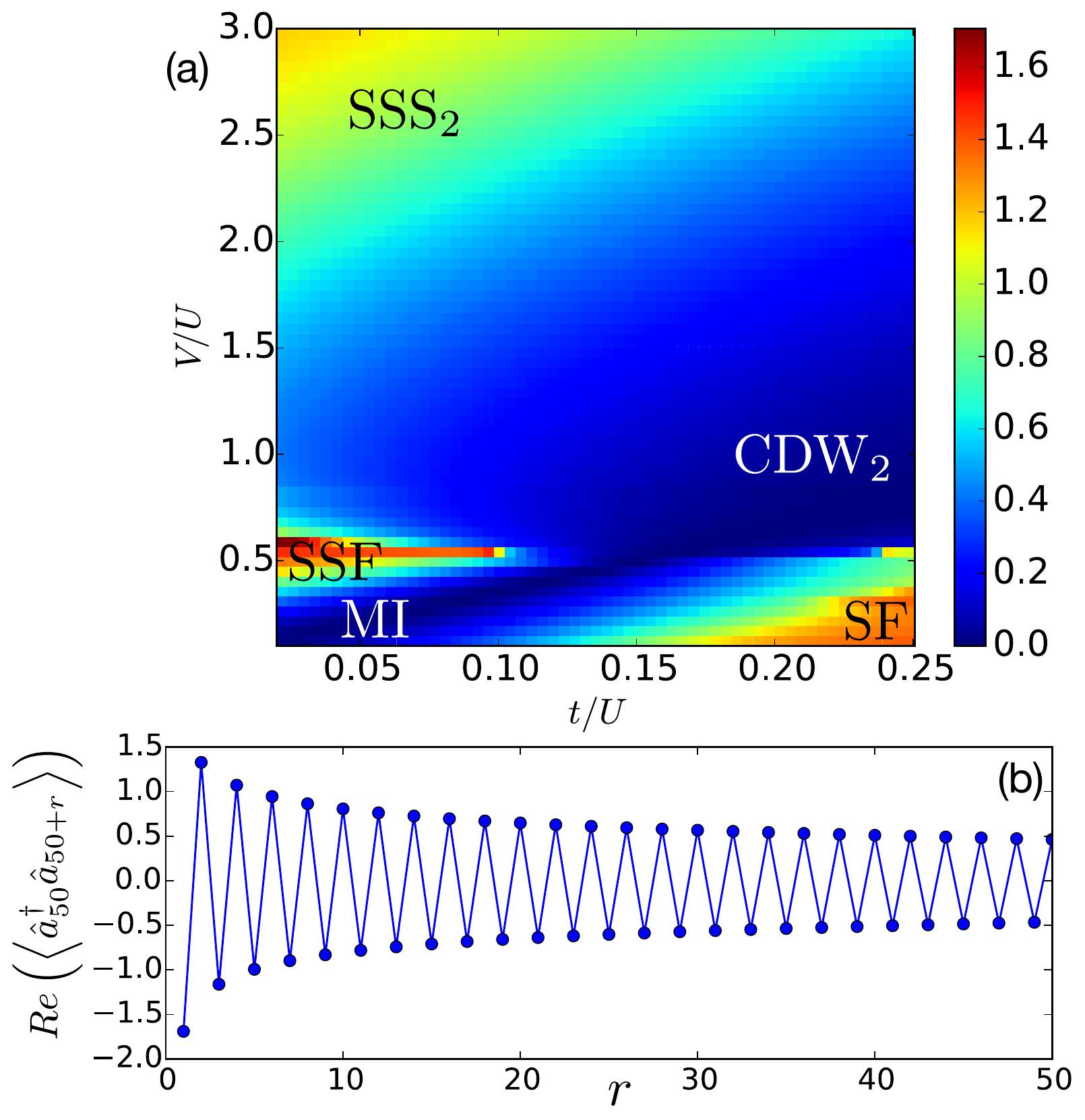}
\caption{(Top panel) Rich phase diagram of dipolar gas in 1D optical lattice for density $\rho=2$ as visualized by the ground state entanglement entropy. The MI and SF phases are accompanied by period-2 charge-density wave (CDW$_2$) and period-2 staggered supersolid (SSS$_2$)
phases. Staggered superfluid (SSF) appears close to $V/U=0.5$. (Bottom panel) The correlation function in the SSF phase in the middle of the chain reveals characteristic oscillations with a power-law decaying envelope. 
The figure has been adapted and reprinted with permissions from~\cite{Kraus20} published in 2022 by the American Physical Society.}
\label{rho2}
\end{figure}

Let us first consider the higher density $\rho=2$. The characteristics of the phases found by MPS-based density-matrix renormalization group (DMRG)~\cite{White1992, White1993} is determined by measuring the site occupation number variance, 
directly proportional to the compressibility 
\cite{Wessel04,Roscilde09,Delande09}, as well as the momentum distribution of the off-diaginal correlations $M(q)$ defined in Eq.~\eqref{eq:Mq} and the structure factor $S(q)$ defined in Eq.~\eqref{eq:Sq}.

The standard superfluid reveals a peak of $M(q)$ at $q=0$, the staggered superfluid (SSF) at $q=\pi$~\cite{Johnstone19}. On the other hand, a maximum of $S(q)$ at $q=\pi$ reveals period-2 density correlations. For an incompressible gapped phase, this will be a period-2 charge-density wave (CDW$_2$) as mentioned earlier, while the compressible gapless phase will be a period-2 supersolid (SS$_2$). When the supersolid phase is accompanied by a staggering in the off-diagonal correlations, i.e., displays a peak of $M(q)$ at $q=\pi$, it will be a period-2 staggered supersolid (SSS$_2$). All these phases are shown in Fig.~\ref{rho2}(a) with the color coding representing the entanglement entropy of the ground state - that is why no border between MI and CDW$_2$ appears in this plot. Figure~\ref{rho2}(b) presents a correlation function of the SSF phase with its staggered shape. Let us note here also that the SSF may also be observed in two-dimensional systems as revealed by cluster mean field study~\cite{Suthar20}. 

\begin{figure}[ht]
	\centering
	\includegraphics[width=\linewidth]{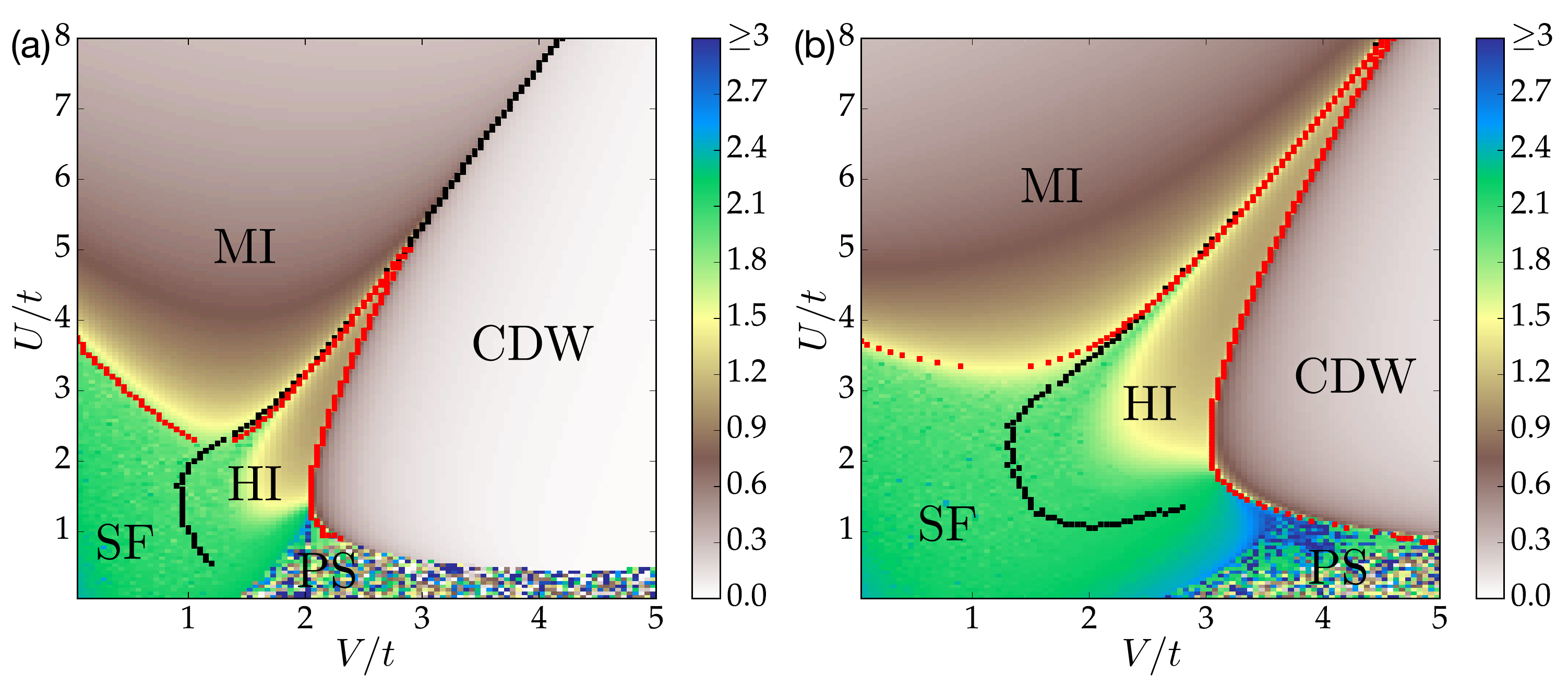} 
	\caption{The phase diagram of EBH at unit filling in the $(U/t,V/t)$-plane obtained with iDMRG. Squares indicate boundaries identified using iDMRG and correspond to the values where the string $\mathcal{O}_{S}(\rho)$ (black) {and/or parity $\mathcal{O}_P$ (red)} order parameters vanish. The left panel shows the von-Neumann entropy for $T\ne 0$, whereas the right panel is for $T=0$. 
    The figure has been adapted and reprinted with permissions from~\cite{Kraus22} published in 2022 by the American Physical Society.} 
	\label{vonNeu}
\end{figure} 

Consider now the often studied unit filling case, mentioned already above, and let us consider the differences in phase diagrams in the presence and in the absence of IIT as shown in Fig.~\ref{vonNeu}~\cite{Kraus20} obtained via infinite DMRG (iDMRG)~\cite{McCulloch2008}. For this comparison, the interactions has been restricted to nearest neighbors only as in the earlier studies~\cite{Rossini12, Batrouni2013, Batrouni2014}. The colors in the figure indicate the value of the von-Neumann entropy of the ground state calculated by splitting the (infinite in iDMRG numerical treatment) 1D chain in two parts. Observe that in the presence of IIT, insulating phases move towards lower interaction values as IIT partially cancels the kinetic tunneling. The borders of the phases in the plot are calculated using order parameters mentioned in Table~\ref{tab:obs}. This also helps to identify the topological Haldane insulator where the string correlations $\mathcal{O}_S^{\alpha}$, $\alpha = x, z$, remain finite in the thermodynamic limit.

\subsection{{The role of the transverse confinement}}
\label{geo:gr}
Above we have assumed a standard $1/r^3$ decay of the inter-site interactions. This decay may be significantly altered by transverse lattice confinement~\cite{Wall13}. For a tight transverse confinement the decay may be faster leading to an effective decay of $1/r^{\beta_{\text{eff}}}$ with 
$\beta_{\text{eff}}>3$ affecting the location of  transition between different phases in the phase diagram. The problem has been revisited recently
\cite{Korbmacher23t} for a {\it weak} transverse confinement after realizing that such a geometry also strongly affects the dynamics for interacting dipoles (see below and~\cite{Korbmacher23}). 

The inter-site interaction between dipoles located at site 0 and $j$ is 
\begin{equation}
V_{j} = \int d^{3}r \int d^{3}r' \ V(\vec{r}-\vec{r}')|W(\vec{r})|^2|W(\vec{r}-ja\vec{e}_{z})|^2,
\label{Eq2}
\end{equation}
with $a$ being the lattice constant and 
\begin{equation}
V(\vec{r}) = \frac{C_{dd}}{4\pi r^{3}}\left(1-3\cos^2\alpha\right).
\end{equation}
the dipolar interaction.

As discussed in detail in~\cite{Wall13}, the inter-site interaction dependence on the distance may be safely estimated (for sufficiently deep optical lattice) using a Gaussian approximation
(note that such an approach is not justified for the calculation of tunneling amplitudes) for the Wannier functions:
\begin{equation}
W(\mathbf{r})=\frac{e^{-z^{2}/2\ell}}{\sqrt{\sqrt{\pi}\ell}}\frac{e^{-(x^2+y^{2})/2\ell_{\perp}}}{\sqrt{\pi}\ell_{\perp}}    
\end{equation}
 with $\ell_{\perp} = \sqrt{\hbar/m\omega_{\perp}}$ being the transverse harmonic oscillator length and $\ell=a/(\pi s^{1/4})$, where $s$ is the depth of optical lattice potential in the units of the recoil energy,  $E_{R}=\frac{\pi^{2}\hbar^{2}}{2ma^{2}}$. With this notation one arrives at~\cite{Sinha07,Deuretzbacher10,Bartolo13,Korbmacher23} 
\begin{equation}
\frac{V_{j}}{E_{R}} = \frac{3B^{3/2}}{2\pi^{2}}(3\cos^{2}\alpha-1)\left(\frac{a_{dd}}{a}\right)f(\sqrt{B}j),
\label{Eq3}
\end{equation}
where $a_{dd} = mC_{dd}/(12\pi\hbar^{2})$ is the dipolar length, $B=\frac{\pi^{2}}{2}\frac{\chi}{1-\frac{\chi}{2\sqrt{2}}}$, $\chi=\hbar\omega_\perp/E_R$, and
\begin{equation}
f(\xi) = 2\xi -\sqrt{2\pi}(1+\xi^2)e^{\xi^2/2}\text{erfc}(\xi/\sqrt{2}).
\label{Eq4}
\end{equation}

Observe that $V_j=VG_j(B)$ where $V=V_1$ and $G_{j}(B) = f(\sqrt{B}j)/f(\sqrt{B})$ depends on the confinement geometry only. Thus fixing the ratio of next-nearest neighbor (NNN) to nearest-neighbor (NN) coupling $V_2/V=1/2^{\beta_{\text{eff}}}$ determines the potential shape. In particular, the ground state
properties are determined by $V/t$ 
and $\beta_{\text{eff}}$. For tight transversal binding, $\ell_\perp << \ell$, $\beta_{\text{eff}}$ may slightly exceed the standard value of 3 corresponding to $\ell_\perp=\ell$. Importantly, for a shallow perpendicular trap $\beta_{\text{eff}}$ may reach much smaller values, strongly affecting the ground state properties.

\begin{figure}
	\centering
	\includegraphics[width=0.85\linewidth]{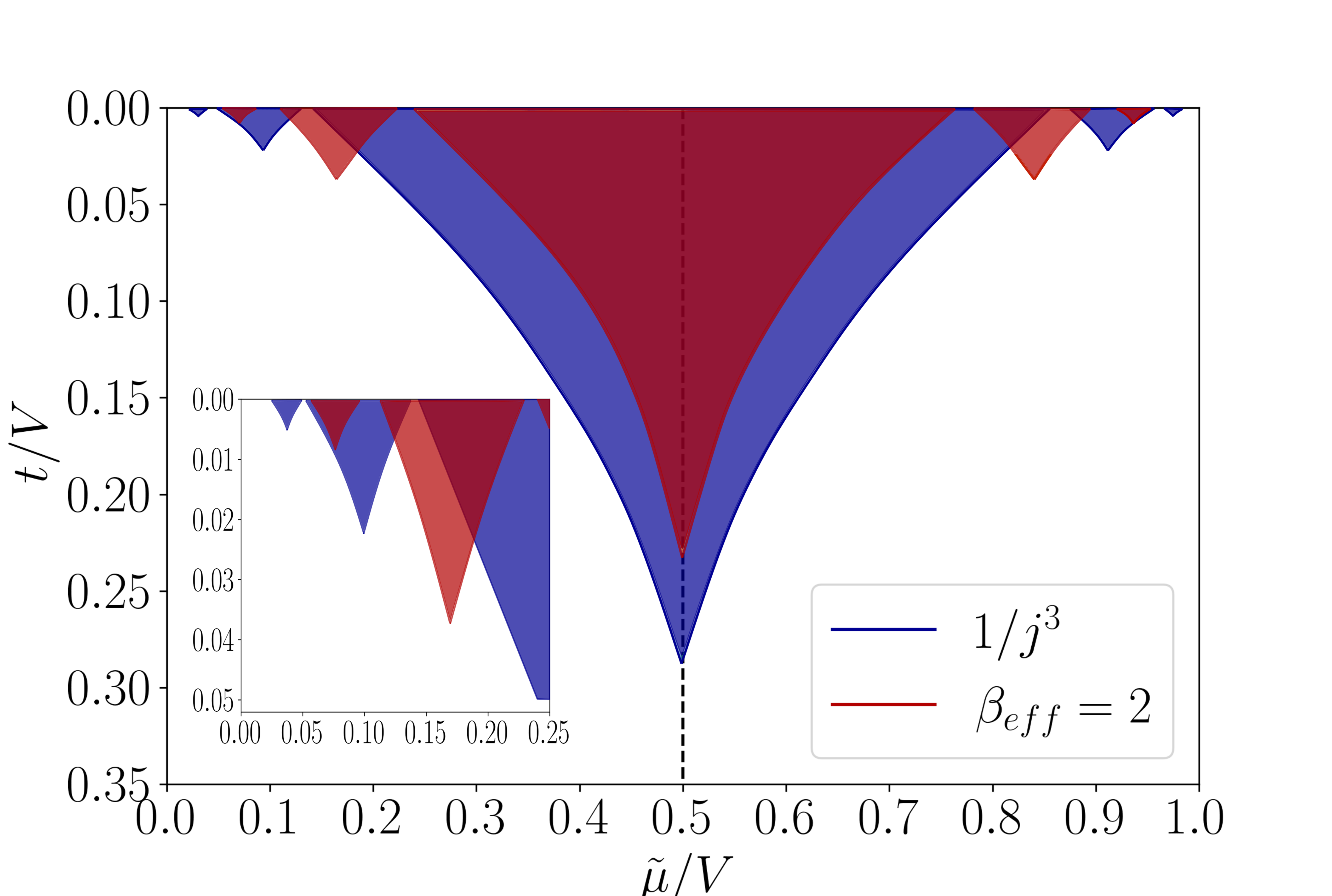} 
	\caption{
Phase diagram in the $(t/V,\tilde\mu/V )$ plane for the standard $\beta_{\text{eff}}=3$ decay (blue lobes) and the modified dipolar interaction with
$\beta_{\text{eff}} = 2$ (red lobes).  The inset reveals the details of the top left corner of the main plot.
 The figure adapted from~\cite{Korbmacher23t} published in 2023 by the American Physical Society.}
	\label{fig:fract}
\end{figure} 

This has been discussed in detail for hard-core bosons in~\cite{Korbmacher23t} both for repulsive and attractive interactions. In the former case, significant shifts of the boundaries of different insulating devil's staircase phases for fractional filling was found - see Fig.~\ref{fig:fract}. The plot utilizes particle hole symmetry of the diagram when represented with the rescaled chemical potential $\tilde\mu=\mu/[2\sum_j G_j(B)]$. For attractive interactions, the standard model predicts the appearance of self-bound lattice droplets~\cite{Morera20,Morera21,Morera23}. Smaller $\beta_{\text{eff}}<3$ interactions result in a reduction of the critical dipole interaction strength for the formation of self-bound clusters, and for an enhancement of the region of liquefied lattice droplets~\cite{Korbmacher23t}.

\begin{figure}
	\centering
	\includegraphics[width=0.99\linewidth]{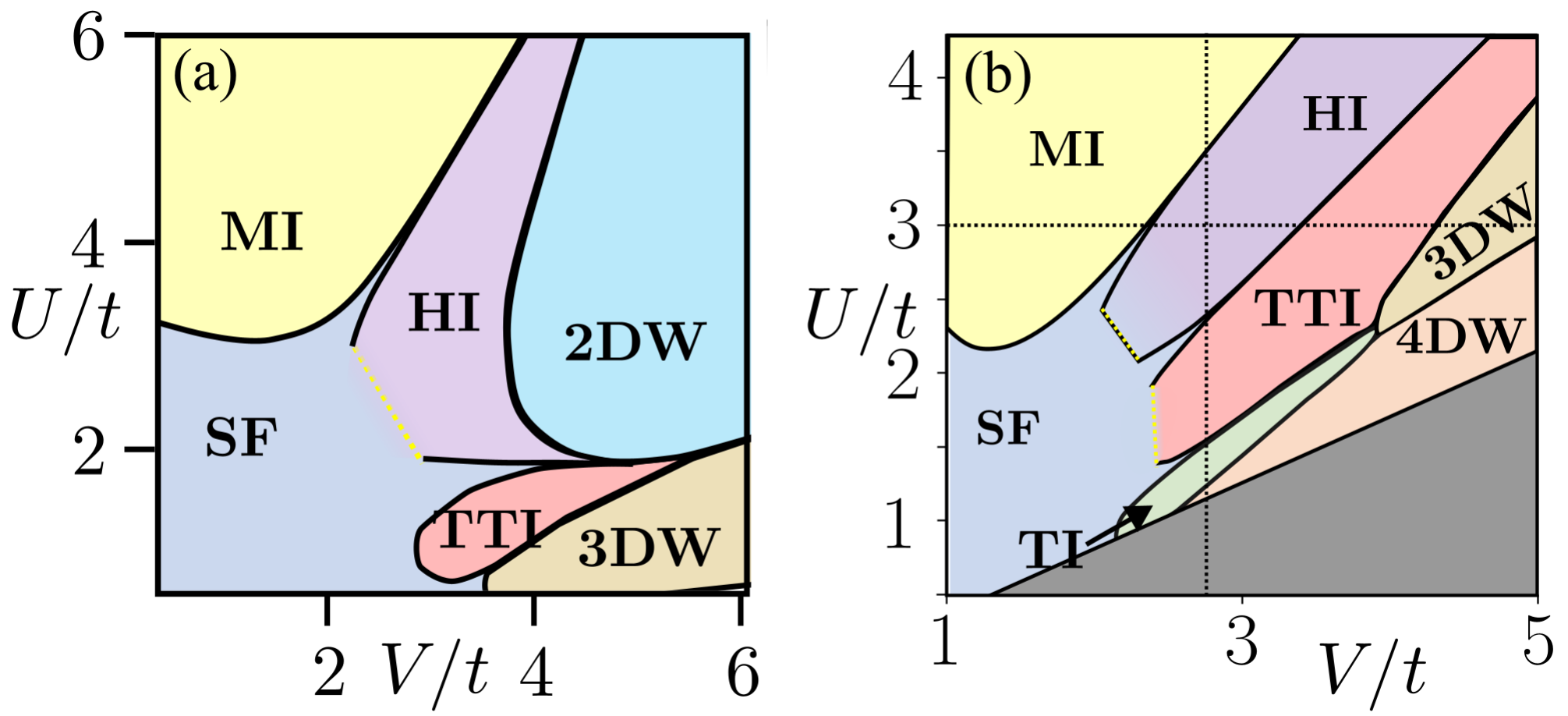} 
	\caption{The phase diagram of EBH at unit filling in the $(U/t,V/t)$-plane obtained with iDMRG for different lattice geometry: (a) Standard dipolar tail for $\beta_{\text{eff}}=3$. Observe the existence of topologically trivial insulator (TTI) absent for the nearest neighbor model; (b) The phases for a shallow perpendicular binding with $\beta_{\text{eff}}=1$. In addition to TTI a new  topological insulator (TI)  phase appears. See text for discussion and~\cite{Lacki23} for more details. The figure has been adapted  from~\cite{Lacki23} published in 2024 by the American Physical Society.} 
	\label{fig:eBH1D}
\end{figure} 

A very interesting case is the celebrated unit density filling in 1D, studied in detail earlier~\cite{Torre06,Deng2011,Dalmonte2011,Rossini12,Batrouni2013,Ejima2015}, as mentioned in Section ~\ref{den-den}. The recently obtained~\cite{Lacki23} phase diagrams are presented in Fig.~\ref{fig:eBH1D} for sufficiently deep lattice ($s=13$) such that the density-dependent tunnelings do not play any role.  In order to identify different density wave insulators, not only CDW$_2$ with period two, but also CDW$_3$ and period-4 CDW$_4$, the iDMRG with 12-sites unit cell was used (assuring convergence with respect to on-site Hilbert space dimension). The dashed yellow lines in the diagrams show that the transition points are not determined accurately in the approach assumed, mainly in the transition between superfluid and Haldane insulator. For small $U$ and large $V$ a region with phase separation is predicted for the EBH model~\cite{Batrouni2013,Batrouni2014,Kottmann2021}, but let us note that this parameter region may be hard to reach experimentally.

 We shall concentrate here, however, on the role of the long-range interaction tail, typically neglected in the standard EBH model. The left panel of Fig.~\ref{fig:eBH1D} depicts the phase diagram for the interactions not limited to nearest neighbors as before, but when the full dipolar tail is taken into account (in numerics, converged results are obtained for interaction ranges bigger than 10 sites). While the occurrence of a period-3 CDW$_3$ -- denoted as a brown patch in the right down corner of Fig.~\ref{fig:eBH1D}(a) -- is to be expected for sufficiently large $V$, a new phase denoted as TTI (topologically trivial insulator)~\cite{Lacki23} also appears in Fig.~\ref{fig:eBH1D}(a). We shall discuss the properties of this phase in detail below. Upon increasing the role of the dipolar tail by making the trap shallow in the directions perpendicular to the one of the lattice, the phase diagram becomes even more interesting as visualized in Fig.~\ref{fig:eBH1D}(b) for $\beta_{\text{eff}}=1$. In addition to different charge density waves and the TTI phase a novel topological insulator is found (denoted as TI), sandwiched between the TTI and CDW$_4$ phase. The gray triangle in Fig.~\ref{fig:eBH1D}(b) is excluded from the calculations. It may contain higher order density waves as well as phase separation as for the standard EBH model~\cite{Batrouni2013,Batrouni2014,Kottmann2021}.

 Let us now discuss the properties of the novel phases found: TTI and TI. Their existence is surprising, since it breaks the common lore (coming from the physics of $S=1$ systems) that the phases at unit density are either the disordered one (i.e., MI), SSB phases (charge density waves) or the topological HI.
 \begin{figure*}%
\includegraphics[width=2\columnwidth]{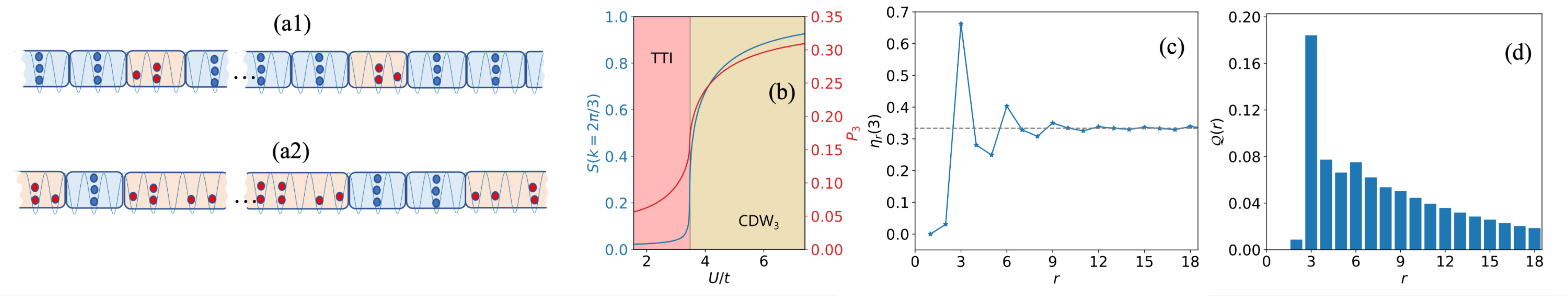}
\caption{(a1) Illustration of the number distribution deep within the CDW$_3$ phase: $(030)$ blocks~(blue slabs) with occasional interspersed defects $(021)$ or $(120)$  (pink slabs) induced by hopping. (a2) In the TTI, regions with empty sites, singlons, and doublons~(pink slabs) are mixed with rare $(030)$ blocks~(blue slabs), which reveal a tendency to form pairs.
(b) Structure factor $S(k=2\pi/3)$ and triplon population $P_3$ at the TTI-CDW$_3$ interface for $U/V=0.74$. 
(c) Triplon correlation $\eta_r(3)$ and (d) 
probability $\mathcal{Q}_r$ to find nearest triplons at a distance $r$~(see text), for $U/t=2.8$ and $V/t=3.8$, deep within the TTI phase. In all cases we consider $\beta_{\mathrm{eff}}=1$. The figure has been adapted  from~\cite{Lacki23} published in 2024 by the American Physical Society.}
\label{fig:3}
\end{figure*} 

The insulating character of the TTI phase is evident from the correlator $C_j(r)$ (see Eq.~\eqref{corr}) exponentially decaying with distance. The entanglement spectrum is not fully degenerate in TTI, ruling out its topological character. Also, contrary to CDW$_2$T and CDW$_3$, the phases that show a pronounced peak in $S(k)$ for $k=\pi/2$ and $k=2\pi/3$, respectively, $S(k)$ does not indicate any spacial periodicity in the TTI phase. TTI is, however, not fully disordered as the MI phase is. Instead, it is characterized by an intricate correlation between occupations of sites, which is not revealed by the standard string correlator \eqref{eq:string}. Let us define an operator~\cite{Lacki23}
 $$\hat{P}_i(q)=\prod_{k\neq s}(\hat{n}-k)/\prod_{k'\neq q}(q-k')$$
which projects into states with $q$ particles at site $i$. Using it we define a generalized string correlator
\begin{equation} R(q)=\lim_{j\to\infty}\frac{\langle\hat{P}_0(q)e^{i\pi\sum_{0<k<j}\sum_s\hat{P}_k(2s)}\hat{P}_j(q)\rangle}{\langle \hat{P}_0(q)\hat{P}_j(q)\rangle},
\label{Rcorrelator}
\end{equation}
which measures the parity of the number of evenly occupied sites between two sites occupied by $q$ particles. Further we define $R_E=\sum_s R(2s)$ and $R_O=\sum_s R(2s+1)$. Let us see how these correlators signal different phases. For the HI we have $R_O\simeq 0$ and $R_E<0$ due to singly occupied sites (singlons) with occasional alternating appearance of empty and doubly occupied sites (doublons). On the other hand, deep in the CDW$_3$ phase, the ground state consists (approximately) of triply-occupied sites~(triplons) $|\dots 030030030 \dots\rangle$, with maximal triplon population $P_3\equiv\sum_i \langle \hat P_i(3) \rangle = L/3$ (with $L$ being the system size). $(030)$ blocks sometimes decay into defects $(120)$ or $(021)$ due to tunneling as shown in Fig.~\ref{fig:3}(a1). Thus for CDW$_3$ phase $R_O>0$, since between two triplons there is always an even number of doublons plus zero occupied sites. Also $R_E=0$ as two doublons or two zero occupied sites are separated by an arbitrary number of evenly-occupied sites. Now assume lowering $V$ along a horizontal line in Fig.~\ref{fig:eBH1D}. The number of defects in the CDW$_3$ phase increases and the system enters the TTI phase roughly when $P_3\simeq L/6$. Interesting insight is further obtained using the parity order associated to pairs $(12)$ or $(21)$, $\mathcal{O}_P(1,2)=\langle (-1)^{\sum_{i<l<j}(P_1(l)+P_2(l))}\rangle$. It is finite inside the CDW$_3$ phase, vanishes at the CDW$_3$-TTI transition and remains zero in the TTI phase where the number of ``defects'' $(021)$ an $(120)$ is strongly increased. Still, it seems that hopping mainly affects doublons, singlons and empty sites located between $(030)$ blocks~(orange region in Fig.~\ref{fig:3}~(a2)). That does not modify $R_E$ or $R_O$, so the correlations $R_E\simeq 0$ and $R_O>0$ are preserved in the TTI phase.

The long-range periodic positional correlations are lost in the TTI phase (as revealed by lack of pronounced peaks in $S(k)$), still short-range correlations do remain. To unravel them, consider $\eta_r(3)=\frac{1}{3}\langle\hat{P}_i(3)\hat{P}_{i+r}(3)\rangle/ \langle\hat{P}_i(3)\rangle^2$, describing the correlation of triplons. In the CDW$_3$ phase $\eta_{3n}(3)=1$ and $\eta_{3n\pm 1}(3)=0$. The triplon correlation $\eta_{3n}(3)$ remains nonzero in TTI phase as presented in Fig.~\ref{fig:3}(c). We may also consider the distribution of triplons separated by $r$ sites
\begin{equation}
 \bar{\mathcal{Q}}_{r}=\langle\hat{P}_j(3)[\Pi_{j<k<j+r}(1-\hat{P}_k(3))]\hat{P}_{j+r}(3)\rangle,   
\end{equation}
which obviously reveals a single peak at $r=3$ deep in the CDW$_3$ phase. In the TTI phase, triplons still like to be $r=3$ apart, for longer distances the distribution resembles poissonian (i.e. random) statistics -- c.f. Fig.~\ref{fig:3}(d). All those discussed residual correlations disappear at the TTI-SF transition.

Consider now the TI phase. Similarly to TTI it is insulating with exponentially decaying correlation $C_j(r)$. $S(k)$ does not reveal any significant positional ordering, in particular the $k=\pi/2$ peak, characterizing CDW$_4$, vanishes at the transition between CDW$_4$ and TI. The standard string order vanishes in this phase. This is due to the fact that both triplons and 4-bosons occupancies are abundant in that phase. While deep in the CDW$_4$ phase, one has $R_E<0$ and $R_O=0$ as appropriate for a repeating $(0400)$ string (possible defects $13$ or $31$ do not affect that). $R_E<0$ persists in the TI phase with $R_O\simeq 0$. Similar behavior is observed for the HI. Importantly, the TI phase satisfies the necessary condition for being a topological phase with a doubly degenerate entanglement spectrum.  While~\cite{Lacki23} does not indicate which symmetries protect the possible topology, a subsequent study~\cite{Lacki24} indicates that the TI is protected by the same lattice inversion symmetry as the HI phase is~\cite{Ejima2014}.

A more detailed analysis reveals~\cite{Lacki23} that HI-TTI and TTI-TI transitions belong to the Luttinger liquid universality class (as the MI-HI transition~\cite{Berg2008, Ejima2014}) with the central charge $c=1$ as obtained from the scaling of the entanglement entropy at the criticality. Similar analysis shows that the HI-CDW$_2$ transition belongs to the Ising universality class with $c=1/2$ in agreement with the finding for extended Hubbard model with nearest neighbour interactions~\cite{Berg2008,Ejima2014}. However, TTI-CDW$_3$ belongs to 3-state Potts universality class of with $c=4/5$ while TI-CDW$_4$ yields again $c=1$ which is attributed to 4-state Potts universality class~\cite{Dotsenko2020}.
 
\subsection{Induced density dependent tunnelings}
\label{induced}

Up till now we considered EBH models with inter-site terms appearing due to interactions. On the other hand, one may design models where the IIT's are induced in the system externally, e.g. due to inter-species interactions or due to the so-called Floquet engineering~\cite{Eckardt2017,Weitenberg2021}. We provide here a few examples of such situations, realizing that the list is far from being complete.

One proposition considers the mixture of two types of particles with each type being confined to its own lattice with the spacing between both lattices being $\lambda/4$. Due to interactions between ``a'' and ``b'' particles, the tunneling of say ``a'' particles depends on the presence or absence of the ``b'' species. Such situations were considered e.g. in~\cite{Cuadra18,Cuadra19,Cuadra19i,Cuadra20,stasi2021}, resulting in novel physics. Interestingly, in the seminal paper~\cite{Jaksch98}, where such a configuration was also considered, the IIT terms were omitted. In particular~\cite{Cuadra18} suggested the existence of a bosonic analog of Peierls transition with spontaneously broken translational symmetry of the underlying lattice. This leads to an analogous to the Su-Schrieffer-Heeger model, a topological insulator in the presence of interactions. The phase diagram of the model shows different types of bond order waves and topological solitons~\cite{Cuadra18,Cuadra19,Cuadra19i,Cuadra20}. Similarly~\cite{stasi2021} finds unusual superfluid phases with clustering properties. Those are probably closely linked to staggered superfluid phases discussed above for dipolar systems. Let us note, that the creation of density dependent tunnelings due to interspecies interactions shows a great similarity with the link model for lattice gauge theory implementations~\cite{Wiese13}. We do not want to review this exciting and rapidly developing field further and refer the reader to recent reviews~\cite{Aidelsburger21,Banuls20}.

A second possibility arises due to Floquet engineering. Consider a standard Bose-Hubbard model, given by the first line of \eqref{ham}, with periodically driven onsite interactions $U(t)=U_0+U_1\cos(\omega t)$~\cite{Gong09,Rapp12}. The corresponding effective time-independent Hamiltonian obtained after averaging the rapidly oscillating terms contains the modified tunneling term, leading to:
 \begin{equation}
     H_{\mathrm eff}=-J\sum_{<ij>}\hat b^\dagger_i {\cal J}_0\left(\frac{U_1}{\hbar\omega}(\hat n_i - \hat n_j)\right)\hat b_i +\frac{U_0}{2} \sum_i \hat n_i(\hat n_i-1).
     \label{driven}
 \end{equation}
Note that the density dependent tunnelings remain as the only possible tunneling mechanism in this scenario. The argument of the Bessel function contains the difference between occupations on the nearby sites. In effect one may expect a strong modification of the phase diagram with creation of pair superfluidity. An extension of this model to two types of particles sitting in nearby sites allows for the creation of density-dependent synthetic gauge fields~\cite{Greschner14}. The experimental demonstration of Floquet engineering based on \eqref{driven} is described below in the experimental Section.

\subsection{Two dimensional extended Bose-Hubbard models}
\label{2D}

Extending the analysis to higher dimensional lattices is of particular interest also from the experimental point of view. Here, various novel phases have been predicted, in particular for standard square lattices, using quantum Monte Carlo (QMC) techniques, as reviewed in~\cite{Dutta15}. Since then, a further significant progress has been made with the help of
EBH models for interacting dipoles with different orientation with respect to the plane of the optical lattice. On one side, Gutzwiller mean field cluster calculations have been developed~\cite{Angom19,Suthar20}, on the other a further spectacular progress has been made with QMC techniques~\cite{Zhang15,Zhang18,Zhang21,Zhang22}. In particular~\cite{Zhang22} brings a most recent analysis for dipoles tilted with respect to the lattice plane, taking into account a possible shaping of lattice sites in the transverse direction. The study is done for different fractional fillings at rather strong on-site interactions $U/t=20$. A multitude of different phases were found that depend on the direction of the dipoles, in particular various types of supersolids (e.g., checkerboard, stripe) and solids (checkerboard, stripe, diagonal stripe). 

Interestingly, also a cluster supersolid is found. It is characterized by the formation of horizontal clusters of particles. These clusters further order along a direction at an angle with the horizontal. This arrangement results from the competition between attractive interaction along the $x$ direction which favors a stripe solid structure, and also attractive interaction along the positive diagonal. Another new phase found is dubbed a grain-boundary superfluid, as in it regions with solid order are separated by extended defects—grain boundaries, supporting superfluidity. In another study of a similar system~\cite{Wu2020} a cluster mean field approach is combined with infinite projected entangled-pair tensor network techniques to improve on the phase boundaries.

Let us mention briefly that other lattice geometries are also considered for dipolar interactions~\cite{Angom22,Hughes22}.

\section{Ground-state physics with cavity-mediated interactions}
\label{sec:cav}

\begin{figure}
\includegraphics[width=\linewidth]{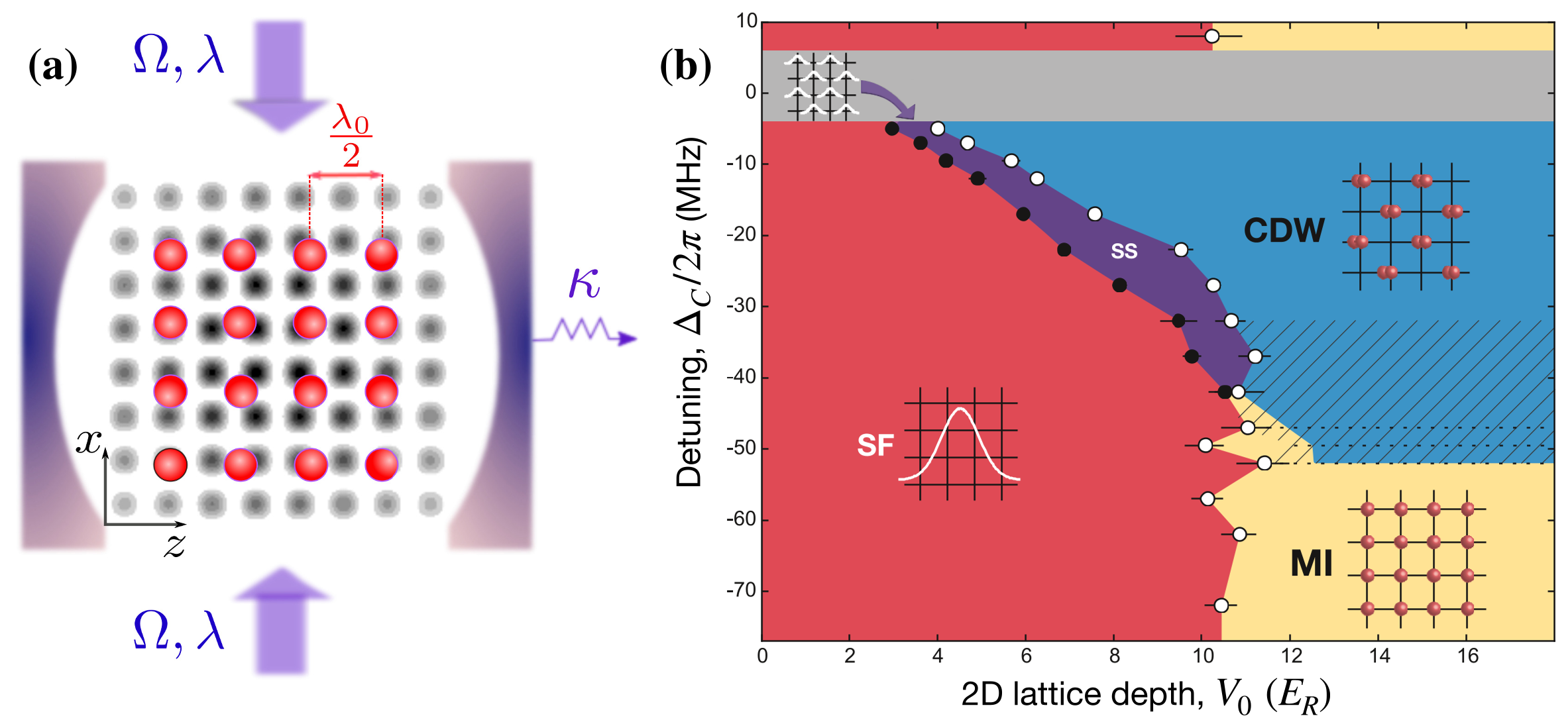}
\caption{(a) Schematic depiction of cavity quantum electrodynamics setup to simulate extended Bose-Hubbard model with long-range interactions in 2D geometry. The ultracold bosons (red spheres) are tightly confined by an optical lattice with lattice constant $\lambda_0/2$ and dispersively interacting with a standing-wave mode of the cavity having wavelength $\lambda$. The setup is driven by a transverse pump laser beams with Rabi frequency $\Omega$. 
Photon scattering off the atoms result into cavity-mediated long-range interactions among the bosons.
(b) The phase diagram of the extended Bose-Hubbard model with cavity-mediated infinite-range interactions experimentally determined by the Quantum Optics Group at ETH Zurich. 
The interplay between the global and short-range interactions gives rise to \textit{superradiant} CDW and SS phases, alongside the standard SF and MI phases.
Panels (a) and (b) have been adapted and reprinted with permissions from~\cite{Habibian2013} published in 2013 by the American Physical Society and~\cite{Landig2016} published in 2016 by the Nature Publishing Group, respectively.
}
\label{fig:cavity_setup_exp}
\end{figure}

Up to this point, our discussion has predominantly revolved around scenarios in which the interaction between the bosons and the trapping laser light is minimal. In essence, this implies that the likelihood of a photon being scattered by a particle is so low that the occurrence of a subsequent scattering event involving the same photon is exceedingly rare. As a result, laser light forms a static ``classical'' optical lattice for the ultracold bosons. However, the dynamics change notably when the optical lattice setup is placed inside a high-finesse optical cavity that is pumped by an external transverse laser field (see Fig. \ref{fig:cavity_setup_exp}(a) for the schematic of the setup in 2D geometry)~\cite{Maschler2005, Maschler2008, FernandezVidal2008, Mekhov2012, Habibian2013, Habibian2013b, Bakhtiari2015, CaballeroBenitez2015, Elliott2016, CaballeroBenitez2016, Dogra2016, Flottat2017, Nagy2018, Chiacchio2018, Himbert2019, Chanda21, Chanda22b, Sharma2022, Fraxanet2023}. In this situation, photons from the pump field get scattered off the atoms and populate the cavity mode(s), and thereby lead to the emergence of effective light-mediated long-range interactions between the bosons due to cavity backaction~\cite{Baumann2010, Habibian2013, Landig2016, CaballeroBenitez2016}.

Such cavity quantum electrodynamics (cQED) setups with bosonic ultracold atoms, recently realized experimentally~\cite{Baumann2010, Klinder2015, Landig2016, Hruby2018, carl2022phases}, have become extremely suitable for realizing quantum simulations of effective many-body long-range Hamiltonians. This has allowed exploration of non-conventional \textit{superradiant}~\cite{Dicke1954, DeVoe1996} quantum many-body phases beyond the typical superfluid and Mott-insulator phases in controllable experimental conditions (see Fig.\ref{fig:cavity_setup_exp}(b) for a recent experimental phase diagram).
It is to be noted that the cQED with ultracold atoms is a rapidly evolving area of research, accompanied by an extensive body of literature. For detailed and comprehensive discussions on this subject, we refer to the recent reviews~\cite{Ritsch2013, Mivehvar2021, Schlawin2022} and references therein.

The typical setup for cQED supporting a single cavity-mode is depicted in Fig.~\ref{fig:cavity_setup_exp}(a) where atoms can move in a 2D layer. Here, a static optical lattice with lattice constant $\lambda_0/2$ (wavenumber $k_0 = 2\pi/\lambda_0$) is placed inside a high-finesse cavity that can support standing-wave modes of periodicity $\lambda_c$ (wavenumber $k_c = 2\pi/\lambda_c$). The setup is driven using a standing-wave pump laser with Rabi frequency $\Omega$, oriented along the $x$-axis. For a large detuning $\Delta_a = \omega_L - \omega_a$ between the pump frequency $\omega_L$ and atomic transition frequency $\omega_a$, excited atomic states can be adiabatically eliminated and the single-particle Hamiltonian in the reference frame rotating at frequency $\omega_L$ is given by~\cite{Baumann2010, Habibian2013, Landig2016}
\begin{align}
\mathcal{H}_{sp} =& \mathcal{H}^0_{sp} + V_1 \cos^2(k_c x + \phi_x) \nonumber \\
&+ \hbar\left(\Delta_c - U_0 \cos^2(k_c z + \phi_z)\right) \hat{a}^{\dagger}\hat{a} \nonumber \\
&+\hbar \eta (\hat{a} + \hat{a}^{\dagger}) \cos(k_c x + \phi_x) \cos(k_c z + \phi_z),
\label{eq:cavity_sphamil}
\end{align}
where $\mathcal{H}^0_{sp}$ is defined in Eq.~\eqref{eq:Hsp}. As before, for the standard isotropic 2D lattice system, we have $\beta = 1$, while $\beta \gg 1$ corresponds to the 1D case. The second term in the equation above is due to the standing-wave potential of depth $V_1 = \hbar\Omega^2/\Delta_a$ along the $z$ direction created by the pump laser. In the third term operators $\hat{a}$ and $\hat{a}^{\dagger}$ denote the photon annihilation and creation operators, $\Delta_c = \omega_L - \omega_c$ is the detuning between the pump frequency $\omega_L$ and the cavity-mode frequency $\omega_c$, and $U_0 = g_0^2/\Delta_a$ is the dynamical Stark shift of a single maximally coupled atom with $g_0$ being the atom-cavity coupling strength. The last term represents a dynamical square-lattice potential describing the coherent pumping of the cavity field via photon scattering by the atoms. Here, $\eta = g_0 \Omega / \Delta_a$ is the amplitude of scattering of a laser photon into the cavity mode by a single atom. The phases $\phi_{x, z}$ denote the phase differences between the cavity mode and the optical lattice along $x$ and $z$ directions respectively.

In the language of second-quantization, the many-body Hamiltonian reads as 
\begin{equation}
\hat{H}_{\text{cavity}} = \int d\mathbf{r} \ \hat{\Psi}^{\dagger}(\mathbf{r})\left[\mathcal{H}_{sp}  + g \hat{\Psi}^{\dagger}(\mathbf{r}) \hat{\Psi}(\mathbf{r}) -\mu \right] \hat{\Psi}(\mathbf{r}),
\end{equation}
where $\hat{\Psi}(\mathbf{r})$ is the bosonic field operator, $g$ is the contact interaction strength and $\mu$ is the chemical potential. In the lowest-order approximation, one can expand the bosonic field operator in the basis of Wannier functions of the lowest Bloch band as $\hat{\Psi}(\mathbf{r}) = \sum_{\mathbf{j}} W_{\mathbf{j}}(\mathbf{r}) \hat{b}_{\mathbf{j}}$, where $W_{\mathbf{j}}(\mathbf{r})$ is the lowest-band Wannier function localized at site $\mathbf{r}_{\mathbf{j}} = (j_x, j_z) \lambda_0/2$, and $\hat{b}_{\mathbf{j}}$ is the corresponding bosonic annihilation operator. Within this approximation, the system is described by the boson-cavity Hamiltonian~\cite{Maschler2008}:
\begin{align}
\hat{H}_{\text{cavity}} =& - \sum_{\mathbf{j}, \hat{\delta}} t_{\mathbf{j} + \hat{\delta}} (\hat{b}^{\dagger}_{\mathbf{j}} \hat{b}_{\mathbf{j} + \hat{\delta}} + \text{h.c.})
+ \frac{U}{2} \sum_{\mathbf{j}} \hat{n}_{\mathbf{j}}(\hat{n}_{\mathbf{j}} - 1) \nonumber \\
&+ \sum_{\mathbf{j}} \left(\hbar V_1 M^x_{\mathbf{j}} - \mu \right) \hat{n}_{\mathbf{j}}
+ \hbar U_0 \hat{a}^{\dagger}\hat{a} \sum_{\mathbf{j}} M^z_{\mathbf{j}} \hat{n}_{\mathbf{j}} \nonumber \\
&+ \hbar \eta (\hat{a} + \hat{a}^{\dagger}) \sum_{\mathbf{j}} \left[ Z_{\mathbf{j}} \hat{n}_{\mathbf{j}} + \sum_{\hat{\delta}}Y_{\mathbf{j} + \hat{\delta}} (\hat{b}^{\dagger}_{\mathbf{j}} \hat{b}_{\mathbf{j} + \hat{\delta}} + \text{h.c.}) \right] \nonumber \\
& - \hbar \Delta_c \hat{a}^{\dagger}\hat{a}.
\label{eq:cavity_hamil_full}
\end{align}
Here, $t_{\mathbf{j} + \hat{\delta}} = \int d\mathbf{r} \ W_{\mathbf{j}}(\mathbf{r}) \left[\mathcal{H}^0_{sp} + V_1 \cos^2(k_c x + \phi_x) \right]W_{\mathbf{j} + \hat{\delta}}(\mathbf{r})$ describes the nearest-neighbor tunneling amplitude and $U = g \int d\mathbf{r} \ W^4_{\mathbf{j}}(\mathbf{r})$ is the onsite Hubbard interaction. The other coefficients are given by the Wannier function overlap integrals as
\begin{align}
M^{\mu}_{\mathbf{j}} =& \int d\mathbf{r} \ W^2_{\mathbf{j}}(\mathbf{r}) \cos^2(k_c \mu + \phi_{\mu}); \mu = x, z, \nonumber \\
Z_{\mathbf{j}} =& \int d\mathbf{r} \ W^2_{\mathbf{j}}(\mathbf{r}) \cos(k_c x + \phi_x) \cos(k_c z + \phi_z), \nonumber \\
Y_{\mathbf{j} + \hat{\delta}} =& \int d\mathbf{r} \ W_{\mathbf{j}}(\mathbf{r}) \cos(k_c x + \phi_x) \cos(k_c z + \phi_z) W_{\mathbf{j} + \hat{\delta}}(\mathbf{r}).
\label{eq:wannier_overlap}
\end{align}
Here, the terms beyond the nearest-neighbor ones are neglected due to the strong localization of the Wannier functions. Due to the same reason, $Y_{\mathbf{j} + \hat{\delta}} \ll Z_{\mathbf{j}}$, except for very fine-tuned scenarios (see below)~\cite{Chanda22b} and hence the $Y_{\mathbf{j} + \hat{\delta}}$ term can be also dropped from the above Hamiltonian.

The EBH model with cavity mediated infinite-range interactions (cEBH) arises after adiabatically integrating-out the cavity degree of freedom in the limit of large detuning $\Delta_c$ and cavity decay-rate $\kappa$. In this limit, the timescale of the atomic dynamics is much larger compared to that of the photons, and thus the cavity field reaches its steady state very fast~\cite{FernandezVidal2008, Habibian2013, Dogra2016}. Assuming $\Delta_c, \kappa \gg U_0 \sum_{\mathbf{j}} M^z_{\mathbf{j}} \hat{n}_{\mathbf{j}}$, up to second-order in $1/\Delta_c$, we arrive at the cEBH Hamiltonian~\cite{Habibian2013, Dogra2016, Landig2016, Chanda21, Chanda22b}:
\begin{align}
\hat{H}_{\text{cEBH}} =& - \sum_{\mathbf{j}} t_{\mathbf{j} + \hat{\delta}} (\hat{b}^{\dagger}_{\mathbf{j}} \hat{b}_{\mathbf{j} + \hat{\delta}} + \text{h.c.})
+ \frac{U}{2} \sum_{\mathbf{j}} \hat{n}_{\mathbf{j}}(\hat{n}_{\mathbf{j}} - 1) \nonumber \\
&+ \sum_{\mathbf{j}} \left(\hbar V_1 M^x_{\mathbf{j}} - \mu \right) \hat{n}_{\mathbf{j}}
+ \frac{U_1}{L}  \hat{\Theta}^2. 
\label{eq:cebh_hamil}
\end{align}
The last term in Eq.~\eqref{eq:cebh_hamil} describes the cavity-mediated infinite-range interaction with strength $U_1 = 2 \hbar \Delta_c \eta^2 L / (\Delta^2_c + \kappa^2)$, with $L$ the total number of lattice sites, and with
\begin{equation}
\hat{\Theta} = \sum_{\mathbf{j}} \left(Z_{\mathbf{j}} \hat{n}_{\mathbf{j}} + \sum_{\hat{\delta}}Y_{\mathbf{j} + \hat{\delta}} (\hat{b}^{\dagger}_{\mathbf{j}} \hat{b}_{\mathbf{j} + \hat{\delta}} + \text{h.c.}) \right)
\label{eq:cavity_theta}
\end{equation}
being a global operator acting on the bosonic degrees of freedom.

In case of attractive cavity-mediated interactions ($U_1 < 0$), the system may attain non-zero $\Theta = \braket{\hat{\Theta}}$ featuring a modulated spatial profile, while the parameter $\Theta$ vanishes for standard MI and SF phases. The phases with non-vanishing $\Theta$ are the superradiant ones where the steady-state cavity field $\hat{a}_{ss} \propto \hat{\Theta}$ becomes finite. Therefore, the parameter $\Theta$ can serve as an order parameter for spatially modulated superradiant phases. The exact nature of the superradiant phases depends on the ratio $\lambda_c/\lambda_0$ and the phase differences $\phi_{x, z}$, as we discuss below.

\subsection{Charge-density wave and supersolid phases}

In a 2D geometry ($\beta=1$), for commensurate cavity-mediated interactions, i.e.,  $\lambda_c/\lambda_0 = 1$, and zero phase differences between the cavity-mode and the optical lattice, i.e.,  $\phi_{x, z} = 0$, one obtains 
\begin{align}
    & t_{\mathbf{j} + \hat{\delta}} = t = \int d\mathbf{r} \ W_{\mathbf{j}}(\mathbf{r}) \mathcal{H}^0_{sp}W_{\mathbf{j} + \hat{\delta}}(\mathbf{r}), \nonumber \\
    & M^{x}_{\mathbf{j}} = M^x =  \int d\mathbf{r} \ W^2_{\mathbf{j}}(\mathbf{r}) \cos^2(k_0 x), \nonumber \\
    & Z_{\mathbf{j}} = (-1)^{j_x + j_z} Z = \int d\mathbf{r} \ W^2_{\mathbf{j}}(\mathbf{r}) \cos(k_0 x) \cos(k_0 z), \nonumber \\
    & Y_{\mathbf{j} + \hat{\delta}} = 0.
\end{align}
The cEBH Hamiltonian \eqref{eq:cebh_hamil} then simplifies further to 
\begin{align}
\hat{H}_{\text{cEBH}} =& - t \sum_{\mathbf{j}} (\hat{b}^{\dagger}_{\mathbf{j}} \hat{b}_{\mathbf{j} + \hat{\delta}} + \text{h.c.})
+ \frac{U}{2} \sum_{\mathbf{j}} \hat{n}_{\mathbf{j}}(\hat{n}_{\mathbf{j}} - 1) \nonumber \\
&+ \left(\hbar V_1 M^x - \mu \right) \sum_{\mathbf{j}}  \hat{n}_{\mathbf{j}}
+ \frac{Z^2 U_1}{L}  \hat{D}^2,
\label{eq:cebh_hamil_cdw} 
\end{align}
with $\hat{D} = \sum_{\mathbf{j}} (-1)^{j_x + j_z} \hat{n}_{\mathbf{j}}$. The same Hamiltonian can also be obtained in 1D geometry ($\beta \gg 1$). The corresponding phase diagram of the system has been studied extensively, both in 2D and 1D settings, using mean-field analysis, different Monte-Carlo methods, exact diagonalization, and tensor network techniques~\cite{Dogra2016, Niederle2016, CaballeroBenitez2016, Flottat2017, Himbert2019, Chanda22b}, as well as in experiments (see Fig.~\ref{fig:cavity_setup_exp}(b)). 

\begin{figure}
\includegraphics[width=\linewidth]{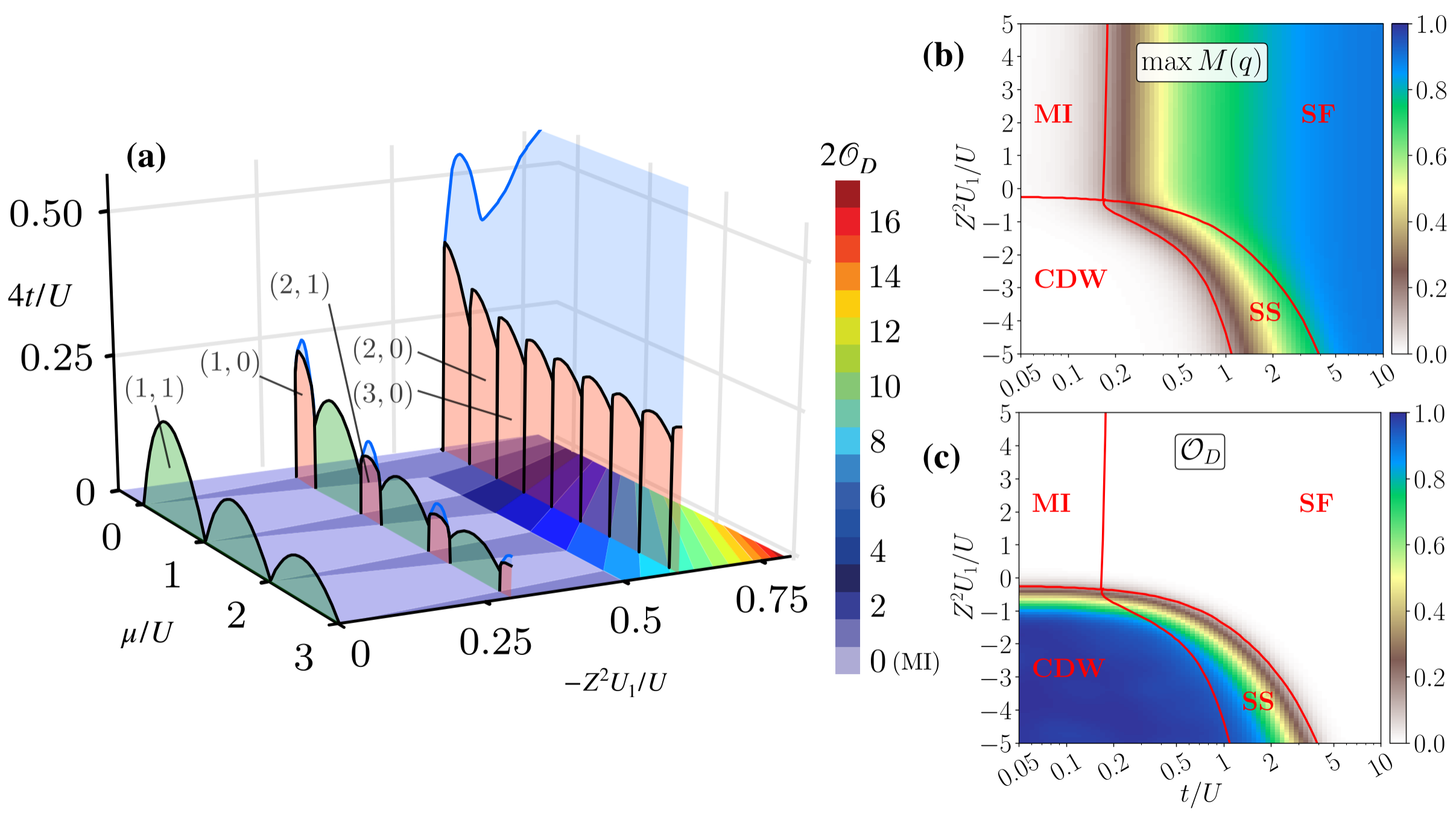}
\caption{(a) The phase diagram of the EBH Hamiltonian \eqref{eq:cebh_hamil_cdw} with cavity-mediated long-range interaction in 2D geometry. The colorbar corresponds $t=0$ plane and indicates (twice) the order parameter $\mathcal{O}_D$ indicating average density imbalance. The transparent colors show the different phases: MI (green), CDW (red), and SS (blue). SF phase is not indicated by colors, but fills the remaining space. The panel has been adapted and reprinted with permissions from~\cite{Dogra2016} published in 2016 by the American Physical Society. (b)-(c) The phase diagram of the same system but in 1D and at unit density. The panels (b) and (c) depict the order parameters $\max_q M(q) = M(0)$ and $\mathcal{O}_D$ respectively. These two panels have been adapted and reprinted with permissions from~\cite{Chanda22b} published in 2022 by the American Physical Society.
}
\label{fig:cavity_cdw}
\end{figure}

In the absence of cavity-mediated interactions, the system refers to the standard BHM, where for integer densities and below a critical hopping amplitude $t < t_c$ the system is in a gapped MI phase, otherwise a (gapless) compressible SF phase is observed. These two phases are distinguished by either the superfluid order parameter $b_{SF} = \braket{\hat{b}}_{\text{avg}}$ or the momentum distribution $M(q) = \frac{1}{L^2} \sum_{\mathbf{j}_1, \mathbf{j}_2} e^{i \mathbf{q}.(\mathbf{j}_1-\mathbf{j_2})} \braket{\hat{b}^{\dagger}_{\mathbf{j}_1} \hat{b}_{\mathbf{j}_2}}$ defined in Eq.~\eqref{eq:Mq} for the 1D scenario. In the SF phase, $b_{SF}$ and $M(0, 0)$ (or $M(0)$ in case of 1D) are finite, while they vanish for the MI phase. In the case of negative cavity interactions, the $\hat{D}^2$ term may favor population imbalance between odd and even sites that spontaneously breaks the discrete $\mathbb{Z}_2$ lattice translational symmetry. For low values of the tunneling amplitude $t$, such $\mathbb{Z}_2$ symmetry breaking results in a CDW phase. This incompressible phase with diagonal (density) long-range order is characterized by finite $\mathcal{O}_D = \frac{1}{L}|\braket{D}|$ and vanishing $M(0, 0)$ (or $M(0)$). In between the gapless SF and gapped CDW phase, an exotic $\mathbb{Z}_2$-broken gapless phase -- the SS phase -- appears, where the diagonal (density) long-range order due to spontaneous breaking of the lattice translational symmetry coexists with the superfluid order. In the SS phase, both $\mathcal{O}_D$ and $M(0, 0)$ are non-zero. The phase diagram of the system is depicted in Fig.~\ref{fig:cavity_cdw}.

It is important to note that here we are ignoring the effect of dipolar interactions among the bosons. Nonetheless, in case of atoms with substantial dipole moments, such as Er or Dy, inter-atomic dipole-dipole interactions with a power-law tail may coexist with cavity-meditated infinite-range density-density interactions. In a recent study~\cite{Hebib2023}, such a scenario has been considered where bosons interact via repulssive $V_{|i-j|} \propto 1/|i-j|^3$ dipolar interactions in 2D setting. Using QMC simulations, it has been shown that CDW and SS phases having checkerboard order get enhanced due to additional inter-atomic dipolar interactions.

\subsection{Bond order and topology}

In case of a commensurate geometry $\lambda_c / \lambda_0 = 1$, when the phase differences $\phi_{x, z}$ between the cavity mode and the optical lattice are not zero, the parameter $Y_{\mathbf{j} + \hat{\delta}}$ does not vanish. Furthermore, the relation $Y_{\mathbf{j} + \hat{\delta}} \ll Z_{\mathbf{j}}$ is no longer valid for $\phi_{x, z} \approx \pi/2$, and thus $Y_{\mathbf{j} + \hat{\delta}}$ may no longer be neglected and the effective EBH Hamiltonian \eqref{eq:cebh_hamil_cdw} needs to be modified. For 1D chains ($\beta \gg 1$) along the $z$ axis (see Fig.~\ref{fig:cavity_setup_exp}(a)), the effective Hamiltonian reads~\cite{CaballeroBenitez2016}:
\begin{align}
\hat{H}_{\text{cEBH}} =& -t \sum_j (\hat{b}^{\dagger}_j  \hat{b}_{j+1} + \text{H.c.}) + \frac{U}{2} \hat{n}_j (\hat{n}_j-1) \nonumber \\
&+ \frac{U_1}{L} (Z \hat{D} + Y \hat{B})^2,
\label{eq:cebh_hamil_cdw_bow}
\end{align}
 where $\hat{D} = \sum_{j} (-1)^{j} \hat{n}_{j}$, $\hat{B} = \sum_{j} (-1)^{j} (\hat{b}_j^{\dagger} \hat{b}_{j+1} + \text{H.c.})$, and the parameters $Z$ and $Y$ are defined by the Wannier function overlap integrals:
\begin{align}
Z =& \int dz \ W_j^2(z) \cos(k_0 z + \phi_z), \nonumber \\
Y =& \int dz \ W_j(z) \cos(k_0 z + \phi_z) W_{j+1}(z). 
\label{eq:cavity_params_zy}
\end{align}
Here, we have identified $j=j_z$, and assumed $W_{j_x}(x) = \delta(\lambda_0 x / 2)$ for $\beta \gg 1$ and $\phi_x = 0$.

For $\phi_z  = 0$ the parameter $Y$ vanishes and we go back to the previous scenario, whereas it becomes finite for $\phi_z = \pi/2$ while $Z$ is zero~\cite{Chanda21, Chanda22b}. The scenario of $\phi_z = \pi/2$ arises when the 1D optical lattice along $z$ has minima (i.e., the lattice sites) at the nodes of the cavity mode. In this case ($Z=0$), for attractive cavity-mediated interactions $U_1 < 0$, the $\hat{B}^2$ term in Eq.~\eqref{eq:cebh_hamil_cdw_bow} induces global correlated hopping among the bosons, and favors \textit{dimerized} bond order $\mathcal{O}_B = \frac{1}{2 L} |\braket{\hat{B}}| \neq 0$ in the system. Apart from the MI and SF phases, the system supports a bond-ordered superfluid (BSF) phase~\cite{CaballeroBenitez2016,Chanda21,Chanda22b} where the Fourier transform $M(q)$ at quasi-momentum $q=\pm \pi/2$ attains sharp-peaks indicating (quasi-)long-range coherence among the bosons. Moreover, in this compressible fluid, $\mathcal{O}_B$ is also non-zero due to spontaneous breaking of the discrete $\mathbb{Z}_2$ translational symmetry by dimerization in alternating bonds. 

\begin{figure}
\includegraphics[width=\linewidth]{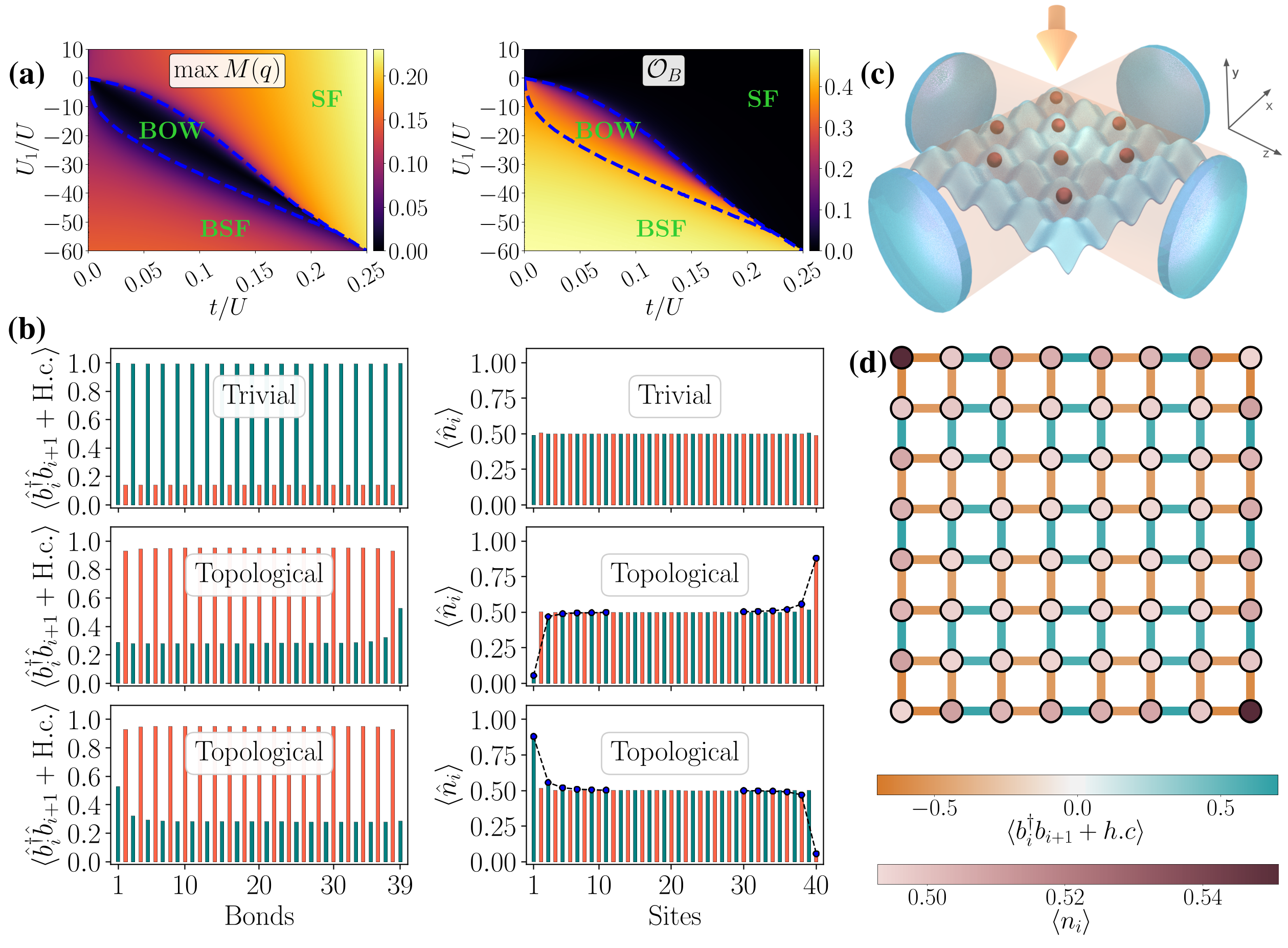}
\caption{Bond-ordered wave and symmetry protected topological phases in the EBH model \eqref{eq:cebh_hamil_cdw_bow} with cavity-mediated long-range interaction at $Z=0$ at half-filling.
(a) The phase diagram of the EBH model in 1D geometry in terms of the maximum of the momentum distribution $M(q)$ and the bond order parameter $\mathcal{O}_B$. The blue dashed lines indicate the borders between different phases. Here, $Y=0.0658$ corresponding to the lattice depth $V_0 = 4 E_R$.
(b) Site-dependent properties of the trivial and topological states of the insulating BOW phase. (Left) Effective tunneling amplitudes $\braket{\hat{b}^{\dagger}_i \hat{b}_{i+1} + {\rm H.c.}}$ as a function of the bonds $(i,i+1)$. Orange (teal) bars denote the even (odd) bond. (Right) Density $\braket{\hat n_i}$ as a function of the lattice site. The dashed lines are guide to the eyes. 
Panels (a) and (b) have been adapted and reprinted with permissions from~\cite{Chanda21} published in 2021 by the Quantum journal.
(c) Two-mode cQED setup for realizing the EBH model of Eq.~\eqref{eq:cebh_hamil_cdw_bow} with $Z=0$ in 2D geometry. Here the atoms are coupled to two cavity modes created by two optical cavities aligned in the $x$ and $z$ directions, and to a laser pump aligned in the $y$ direction. In each direction, the relative phase between the optical lattice (blue) and the cavity mode (orange) is chosen such that the nodes of the latter coincide with the lattice sites.
(d) Topological corner states in the 2D EBH model \eqref{eq:cebh_hamil_cdw_bow} with $Z=0$. The panel shows real-space bond pattern and local occupation for the topological configuration for a system with lattice sites $L=10 \times 10$.
Panels (c) and (d) have been adapted and reprinted with permissions from~\cite{Fraxanet2023} published in 2023 by the American Physical Society.
}
\label{fig:cavity_topo}
\end{figure}

The most interesting scenario occurs for half-integer densities, where an incompressible insulating phase appears between the gapless SF and BSF phases (see Fig.~\ref{fig:cavity_topo}(a)). This insulating phase is a bond-order wave (BOW) with finite $\mathcal{O}_B \neq 0$ and vanishing $M(q) = 0$. The emergence of the BOW phase corresponds to a bosonic Peierls insulator where dimerization by $\mathbb{Z}_2$ symmetry breaking is driven by atom-photon interaction -- reminiscent of the Peierls transition driven by electron-phonon interactions in the Su-Schrieffer-Heeger (SSH) model~\cite{Su1979, Su1980}. Furthermore, similar to the SSH model, the BOW phase is a symmetry protected topological (SPT) phase that is characterized by the existence of two-fold degenerate edge states having particle-hole excitations on the edges (see Fig.~\ref{fig:cavity_topo}(b)), and other indicators of a topological phase, such as non-zero string order, degeneracy in the entanglement spectrum, quantized many-body Berry phase etc.~\cite{Chanda21}. 

It is important to note that in this system the two-fold degenerate topological states are only quasi-degenerate with the non-topological ground state. This is because for finite sizes with open boundaries, the lowest-energy state is the one with positive values of $\braket{\hat{b}^{\dagger}_j \hat{b}_{j+1} + \text{H.c.}}$ at the boundaries. However, the topological states can be reliably prepared by implementing a tailored SSH-like alternating potential and subsequently removing it adiabatically. The two dimensional version of the system \eqref{eq:cebh_hamil_cdw_bow} with $Z=0$ is rich as well~\cite{Fraxanet2023}. In~\cite{Fraxanet2023} an alternate setup for realizing the system with two single-mode cavities has been proposed (Fig.~\ref{fig:cavity_topo}(c)), and a higher order SPT phase has been observed with corner states (Fig.~\ref{fig:cavity_topo}(d)) via bosonic Peierls transition.

In the scenarios where the phase difference $\phi_z$ is not zero or $\pi/2$, e.g., $\phi_z=\pi/4$, both the coefficients $Z$ and $Y$ in Eq.~\eqref{eq:cavity_params_zy} are finite. In such cases, the spatial modulation appears both in the lattice sites and the bonds, and the system supports a gapless bond-ordered supersolid phase, and an insulating phase with both charge-density wave and bond order, in conjunction with the MI, SF, and SS phases~\cite{CaballeroBenitez2016, Chanda22b}.

\begin{figure}
\includegraphics[width=\linewidth]{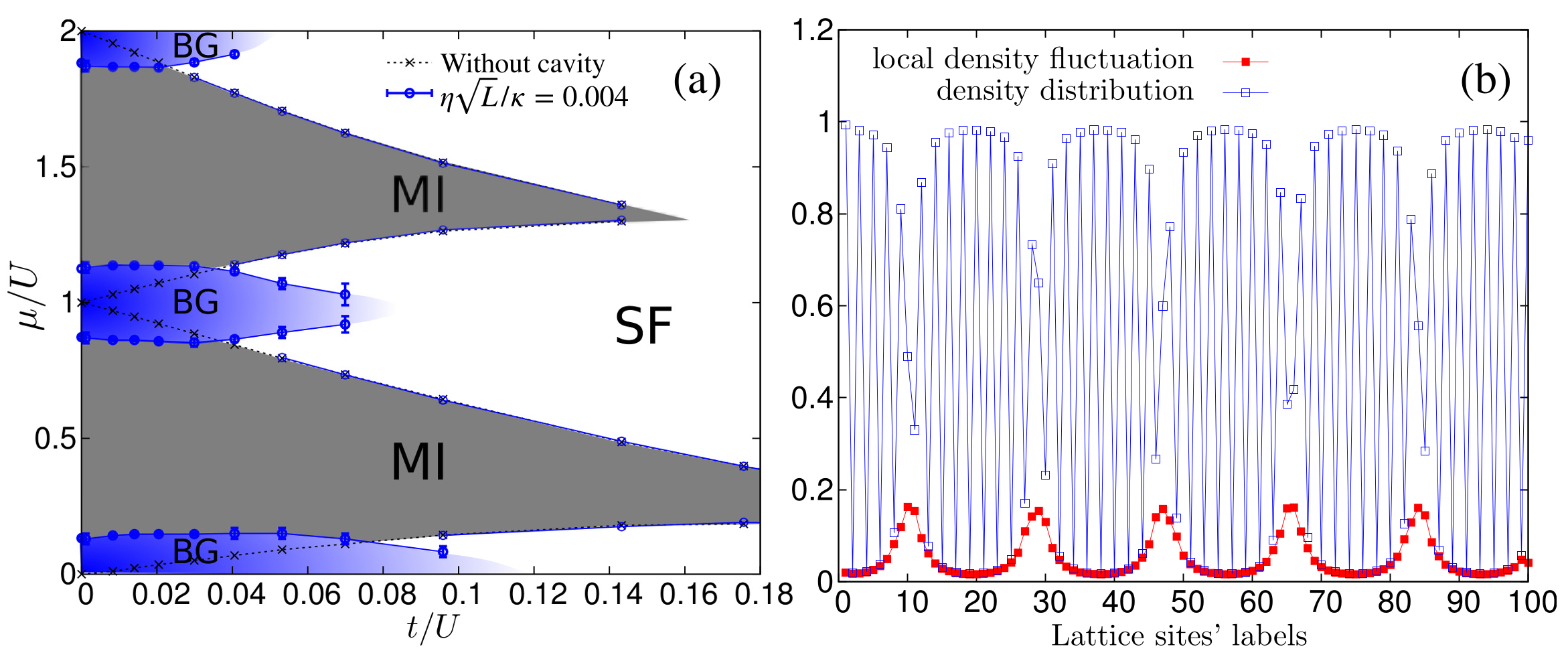}
\caption{(a) The phase diagram of the 1D EBH model with incommensurate cavity-meditated interactions for (semi-)irrational ratio $\lambda_c/\lambda_0 = 830/785$ and $\eta\sqrt{L}/\kappa = 0.004$. The gray regions indicate incompressible MI phases at integer densities and the blue regions indicate gapless and compressible Bose glass (BG) phases with vanishing superfluid order.
(b) Local density $\braket{\hat{n}_j}$ and local density fluctuations $\braket{\hat{n}_j^2} - \braket{\hat{n}_j}^2$  
as a function of the site index in the BG phase for $\mu=0$ and $\eta\sqrt{L}/\kappa = 0.004$.
The panels have been adapted and reprinted with permissions from~\cite{Habibian2013} published in 2013 by the American Physical Society.
}
\label{fig:cavity_bg}
\end{figure}

\subsection{Incommensurate cavity and the Bose glass phase}

Let us now consider the situation when the wavelength $\lambda_c$ of the cavity mode is not commensurate with the optical lattice wavelength $\lambda_0$, i.e., when $\lambda_c/\lambda_0$ is not a rational number (for simplicity, here we fix $\phi_{x, z} = 0$, and assume $\hat{\Theta} = \sum_{\mathbf{j}} Z_{\mathbf{j}} \hat{n}_{\mathbf{j}}$ as usually $Z_{\mathbf{j}} \gg Y_{\mathbf{j} + \hat{\delta}}$).
In such instances, the cavity-mediated interaction induces an effective quasi-periodic potential for the bosons~\cite{Habibian2013, Habibian2013b, Niederle2016} -- a situation that is reminiscent of cold atoms confined in bichromatic quasi-periodic optical lattices~\cite{Deng2008, Roux2008, Deissler2010}. Figure~\ref{fig:cavity_bg}(a) shows the phase diagram of the EBH model with incommensurate cavity in a 1D ($\beta \gg 1$) setup. Apart from standard MI and SF phases, the system supports gapless Bose glass (BG) phases sandwiched between different MI lobes. Since the strength of the cavity interaction now oscillates at a wavelength $\lambda_c$ which is incommensurate with respect to the optical lattice wavelength $\lambda_0$ (see Eqs.~\eqref{eq:cavity_hamil_full}-\eqref{eq:cavity_theta}), the atoms fails to develop proper CDW order due to incommensurability effects. Instead, the system features a gapless compressible state ($\partial \bar{n}/\partial \mu = 0$ with $\bar{n}$ being the average density) with vanishing superfluid order and leading CDW instability, resulting in the BG phase akin to disordered Bose-Hubbard systems. In this phase, the leading CDW instability manifests in quasi-periodic density modulations that oscillate with the beating frequency $|k_{c} - k_0|/2\pi$ (see Fig.~\ref{fig:cavity_bg}(b)). In 1D, the BG phase corresponds to a superradiant phase where $\braket{\hat{\Theta}}$ is non-zero and the cavity field is finitely populated as the atoms coherently scatter the photons coming from the pump field into the cavity at wavelength $\lambda_c$. On the other hand, in 2D (i.e., $\beta=1$), the BG phase can also occur due to the pump field being incommensurate with the optical lattice resulting in a `disordered' chemical potential for the lattice bosons (see the third term in Eq.~\eqref{eq:cavity_hamil_full} or in Eq.~\eqref{eq:cebh_hamil}). In such cases, finite superfluid order can coexists with the CDW leading instability in the BG phases~\cite{Habibian2013b, Niederle2016}.

\section{Excited states dynamics}
\label{sec:dyn}
While up to now we have considered mainly the properties of low lying (or just solely ground) states, interesting physics may occur for highly excited states. Those are often associated with many-body localization (MBL) phenomena that are postulated for strongly disordered systems~\cite{Basko06}. Recent doubts~\cite{Suntajs20e} concerning the existence of MBL in the thermodynamic limit~\cite{Panda20,Sierant20,Abanin21,Sierant20p} made some authors to shift the MBL border to very large disorder values~\cite{Morningstar22,Sels22}, while a search for models different than the paradigmatic Heisenberg chain resulted in studies of the Quantum Sun model~\cite{Suntajs22,Pawlik23,Suntajs23s}, which shows a genuine MBL transition in the thermodynamic limit.

Already much earlier, a similar discussion concerning the existence of MBL in systems with power law decaying tunnelings or interactions took place~\cite{Yao14,Burin15,Burin15l,Li16,Maksymov17,Maksymov20}, and similar studies have been recently performed for infinite-range cavity-induced interactions~\cite{Sierant19,Kubala21,Chanda22c}. The character of MBL changes in such systems with respect to short range models. Without going into details that are beyond the scope of this review let us mention that the thermodynamic limit problem in these cases becomes even more cumbersome. 

On the other hand, typical quantum simulators work in finite size configurations. Additionally, the times at which one may consider such systems to be effectively isolated from the surroundings are finite. From a pragmatic point of view, the thermodynamic limit is then not of primary importance, as cold-atomic systems are typically finite and coherent dynamics may be observed, e.g. in optical lattices, for up to at most several hundreds of tunneling times~\cite{Scherg21}. Thus, a slow approach to ergodicity may be unnoticed - on experimental scales one may still observe spectacular nonergodic dynamics.

Such a situation may happen also in systems without disorder. A seminal example of such a scenario was discussed by Carleo and collaborators~\cite{Carleo12} for the standard Bose-Hubbard model. If the onsite interaction, $U$, between bosons is sufficiently strong, the dynamics may slow down, the spectrum of the BH Hamiltonian becomes fragmented, and a slow nonergodic dynamics may be realized. This is due to repulsively bound pairs (doubly occupied sites or doublons dynamically formed due to an additional energy, $U$, that makes their decay via tunneling costly) whose motion is effectively slowed down ~\cite{Carleo12}. Additional slow down is due to an effective attraction of doublons which makes their clusters hard to break.

Note that too large $U$ may results in an effective splitting of the Hilbert space between states with at most singly occupied sites and those with multiple occupations. The latter, due to high energy cost, are necessarily lying high in energy and affect little the evolution of the singly occupied subspace. The low energy subspace is then well described in terms of hard-core bosons, appearing already several times in this review. 

\subsection{Out-of-equilibrium dipoles}

The similar situation for dipolar particles was analyzed in~\cite{Barbiero2015}. Here the long-range strong interactions may lead to inter-site doublons, i.e. situations, where two atoms on neighboring sites are bound together.  Moreover, clusters of bigger sizes are possible to be formed - those move even slower than inter-site doublons. The effective dynamics becomes extremely slow, resulting in the fact that if some non-equilibrium steady state is formed, it remembers its initial configuration for quite a long time, often exceeding the possible experimental time. In this way a quasi-many-body localization is formed involving those large clusters, realizing localization even in the absence of disorder.

While the dynamics described in~\cite{Barbiero2015} considered very strong dipolar attractive interactions with $V=-100t$, it turns out that nonergodic dynamics appears also for much weaker interactions. As pointed out in~\cite{Li21} it is particularly important to take into account the dipolar tail of the interactions and to not restrict it to the leading nearest neighbor (NN) terms. Due to the, typically, $1/r^3$ tail, the next-nearest neighbor (NNN) interaction strength is just 1/8 of the NN contribution. Taking into account NNN terms may lead, in the spectral domain, to Hilbert space shattering~\cite{Khemani20}, and in the time dynamics to the appearance of the intermediate time scale in the evolution when the dynamics seems not thermal. As long as approximate constants of motion of the system, i.e. the number of NN pairs and the number of NNN pairs, are conserved, the dynamics seems nonergodic, and the time evolved state resembles its initial state. Eventually, approximately conserved quantities are destroyed and the system becomes ergodic at large time scales, however, that may be typically well beyond the capability of current experimental realizations. Thus, for times of the order of hundreds of tunneling times, the dynamics may appear to be effectively strongly nonergodic and apparently many-body localized~\cite{Li21}. Only at later times one observes the decay of the prethermalized phase towards the ergodic long time characteristics. Let us stress that this behavior is observed for hard-core bosons with strong dipole-dipole interactions of $V/t\approx 50$.  

\begin{figure}
\includegraphics[width=\linewidth]{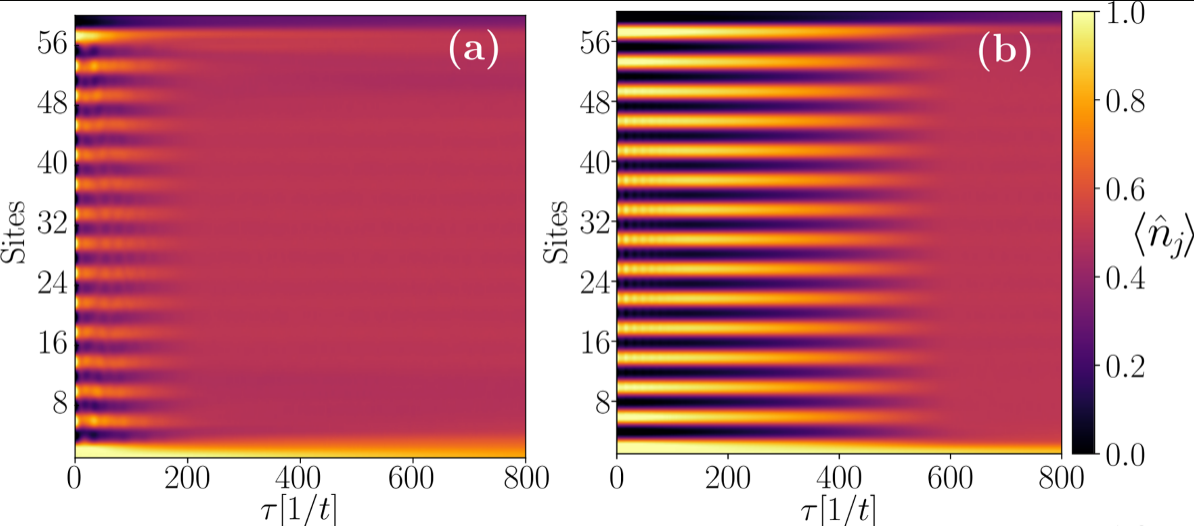}
\caption{Time evolution of the initial density wave~\ref{dw}, for $V/t=16$ and a) $V_j=V/j^3$ and (b) $V_j=VG_j(B)$ with $B=2.54$ ($\beta_{\text{eff}}\approx 2$). The system size is $L=60$ and the result is obtained with time-dependent variational principle algorithm~\cite{Haegeman2011, Haegeman2016, Paeckel2019}.  The has  been adapted and reprinted
with permissions from \cite{Korbmacher23} published in 2023 by the American Physical Society. 
}
\label{fig:dyn1}
\end{figure}

As discussed in Section~\ref{geo:gr}, the inter-site interactions may be significantly altered by changing the transverse confinement. This in turn affects not only the ground state properties as discussed there, but also the time dynamics~\cite{Korbmacher23}. Changing the interactions from $1/2^3$ to $1/2^\beta_{\text{eff}}$ with $\beta_{\text{eff}} < 3$ enhances long-range interactions. One might naively expect more ergodic dynamics in such a case due to the enhanced interactions range. In fact the situation is opposite, small $\beta_{\text{eff}}$ enhances the role of NNN couplings that were found to be particularly important for the Hilbert space shuttering mechanism~\cite{Li21}. This is reflected in the time dynamics as exemplified in Fig.~\ref{fig:dyn1} for the hard-core interacting bosons model. As the initial state the following density wave is taken:
$$
|\fullcirc\,\fullcirc\, \emptycirc\,\emptycirc\,\fullcirc\,\fullcirc\, \emptycirc\,\emptycirc\, \fullcirc\,\fullcirc\, \emptycirc\,\emptycirc\,\cdots\rangle,
\label{dw}
$$
(where $\fullcirc$ ($\emptycirc$) denote filled (empty) site, respectively. Such a state may be prepared experimentally using superlattice techniques~\cite{Guardado21}). Fig.~\ref{fig:dyn1} shows the time dynamics for such an initial state for $V_j=V/j^3$ and for $B=2.54$ ($\beta_{\text{eff}} \approx 2$). In the latter case, the initial state density pattern is preserved for a much longer time, exceeding 500 tunneling times. This correlates well with the result of~\cite{Li21} who found that NNN terms slow down the relaxation. While NNN terms are present in both simulations, they are more significant for the case of smaller $\beta_{\mathrm eff}\approx 2$, explaining the stronger nonergodic effect. As discussed in the original work~\cite{Korbmacher23}, the nonergodic character of the dynamics for strong dipolar interactions is within the experimental reach for $^{164}$Dy in an UV lattice or for NaK molecules in a $\lambda=500\,$nm lattice.

\subsection{The role of density-dependent tunnelings for the motion of soft-core bosons}

While up-till-now we have discussed only the hard-core bosonic systems, it is also interesting to consider the general soft-core bosons and, in particular, the role of interaction-induced tunnelings. As for the ground state properties we consider now the full Hamiltonian of the problem
\begin{eqnarray}
\hat{\mathcal{H}}_{EBH} &=& -t\sum_{j=1}^{L-1} \left (\hat{b}^{\dagger}_j\hat{b}_{j+1}+\mathrm{H.c.} \right )+\frac{U}{2}\sum_{j=1}^L\hat{n}_j(\hat{n}_j-1)\nonumber \\
&+&\frac{V}{2}\sum_{i\neq j}\frac{1}{\vert i-j \vert^3  }\hat{n}_i\hat{n}_j  \nonumber \\
&-&T\sum_{j=1}^{L-1} \left [\hat{b}^{\dagger}_j(\hat{n}_j+\hat{n}_{j+1})\hat{b}_{j+1}+\mathrm{H.c.} \right ].
\label{eq:ext}
\end{eqnarray}
with on-site interaction $U$ (consisting of contact and dipolar terms) as well as interaction-induced tunnelings (IIT) $T$. We fix $U/t=3$~\cite{Adith23} and adjust $V$ and $T$ by changing the lattice depth $s=U_0/E_R$ (this requires tuning the contact interactions via a suitable Feshbach resonance). 

\begin{figure}%
	\includegraphics[width=\linewidth]{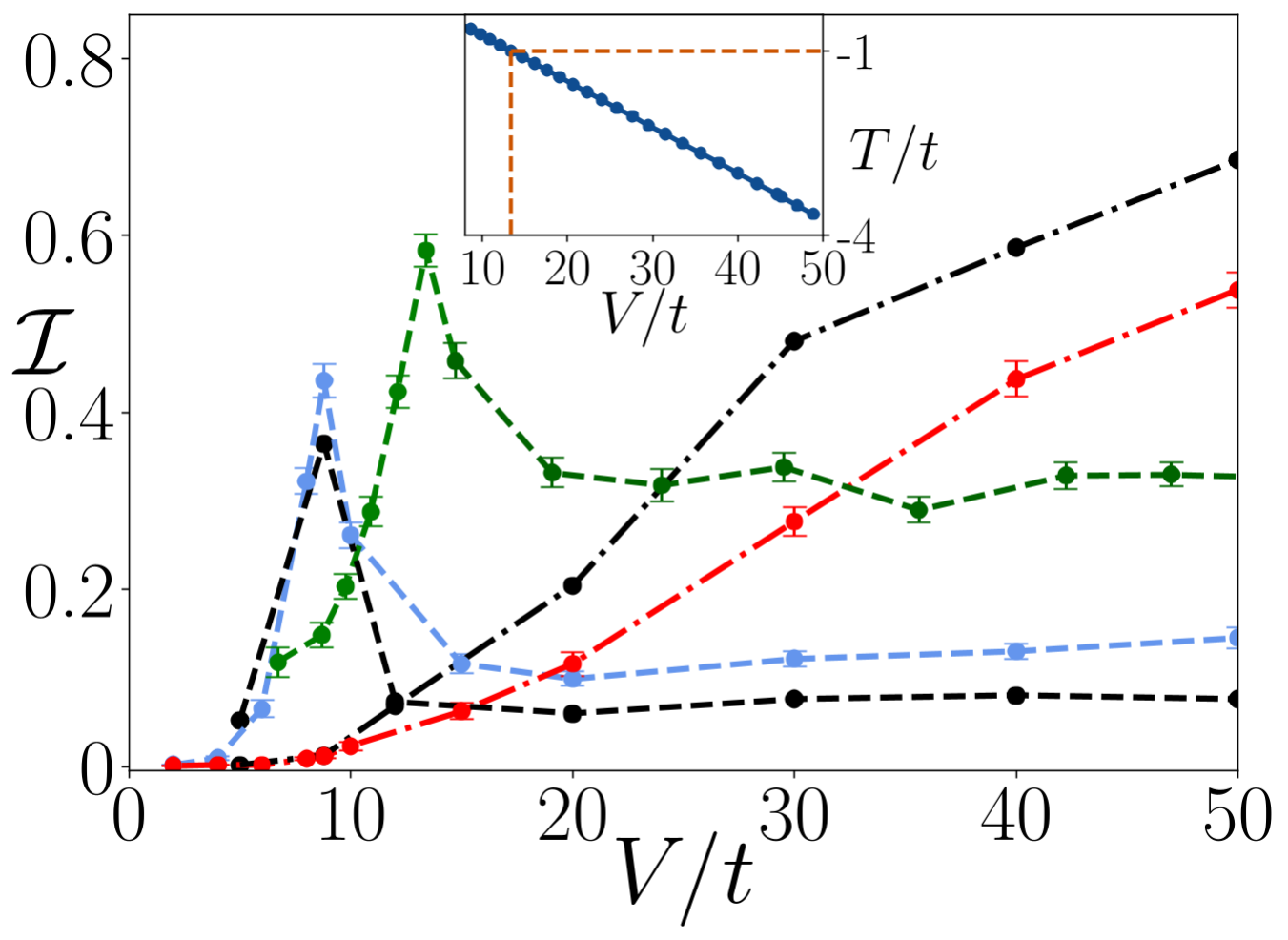}  
	\caption{Final inhomogenity, $\mathcal{I}(\tau_f)$, as a function of $V/t$, for $N=6$, $L=12$. In the absence of IIT (dot-dashed: red~(black) curve for the $N_{\rm NN}=2$~(3) sector), the dynamics becomes steadily more non-ergodic with increasing $V/t$.  The blue~(black) dashed line show our results for $s=8$ 
	for the $N_{NN}=2$~($3$) in the presence of IIT. The green dashed line corresponds to a deeper $s=10$ lattice (for the $N_{NN}=2$ sector). Observe that the peak of enhanced inhomogeneity depends on the lattice depth, being at $V/t=8.8$ for $s=8$ and at $V/t\approx 13$ for $s=10$. The inset shows the dependence of $T/t$ on $V/t$ for $s=10$. The confluence of tunnelings $T/t=-1$ occurs for $V/t\approx 13$.
 The figure has been adapted and reprinted
with permissions from \cite{Adith23} published in 2023 by the American Physical Society.}
	\label{fig:Deloc_zak}
\end{figure} 

Let us consider such a system with half-filling and look at the dynamics. As a measure of the inhomogeneity we take~\cite{Adith23}: 
\begin{eqnarray}
\mathcal{I}(\tau) = \frac{\sum_{i=1}^L\left ( \langle \hat{n}_i(\tau) \rangle - \rho \right )^2}{\sum_{i=1}^L\left( \langle \hat{n}_i(0) \rangle -\rho \right )^2},
\label{eq:zak_fac}
\end{eqnarray}
with $\rho=N/L=1/2$ being the overall particle density. The normalization assures that $0<\mathcal{I}(\tau)<1$, interpolating between being homogeneous and fully correlated with the initial state density. For the latter we assume a random Fock-like separable state with a well defined number of nearby pairs $N_{\rm NN}$, as this determines the dynamics~\cite{Li21} for large $V/t$. 

The final time inhomogeneity, $\mathcal{I}$  from exact time propagation up to time $\tau_f=500/t$ for $N=6$ bosons on $L=12$ sites with open boundary conditions is plotted in Fig.~\ref{fig:Deloc_zak} as a function of $V/t$ in the presence and in the absence of interaction induced tunneling $T$. One can observe that for large $V/t$ the interaction induced tunneling partially restores ergodicity - in fact this process dominates tunneling. On the other hand, there exists an optical lattice depth, $s$, depending on the strength of the interaction value $V/t$, for which $T=-t$, where destructive interference between the kinetic and interaction driven tunnelings occurs [compare \eqref{eq:tuneff}]. This point manifests itself as a spectacular maximum of the inhomogeneity for relatively low $V/t$.

\begin{figure}[t!]
\includegraphics[width=\linewidth]{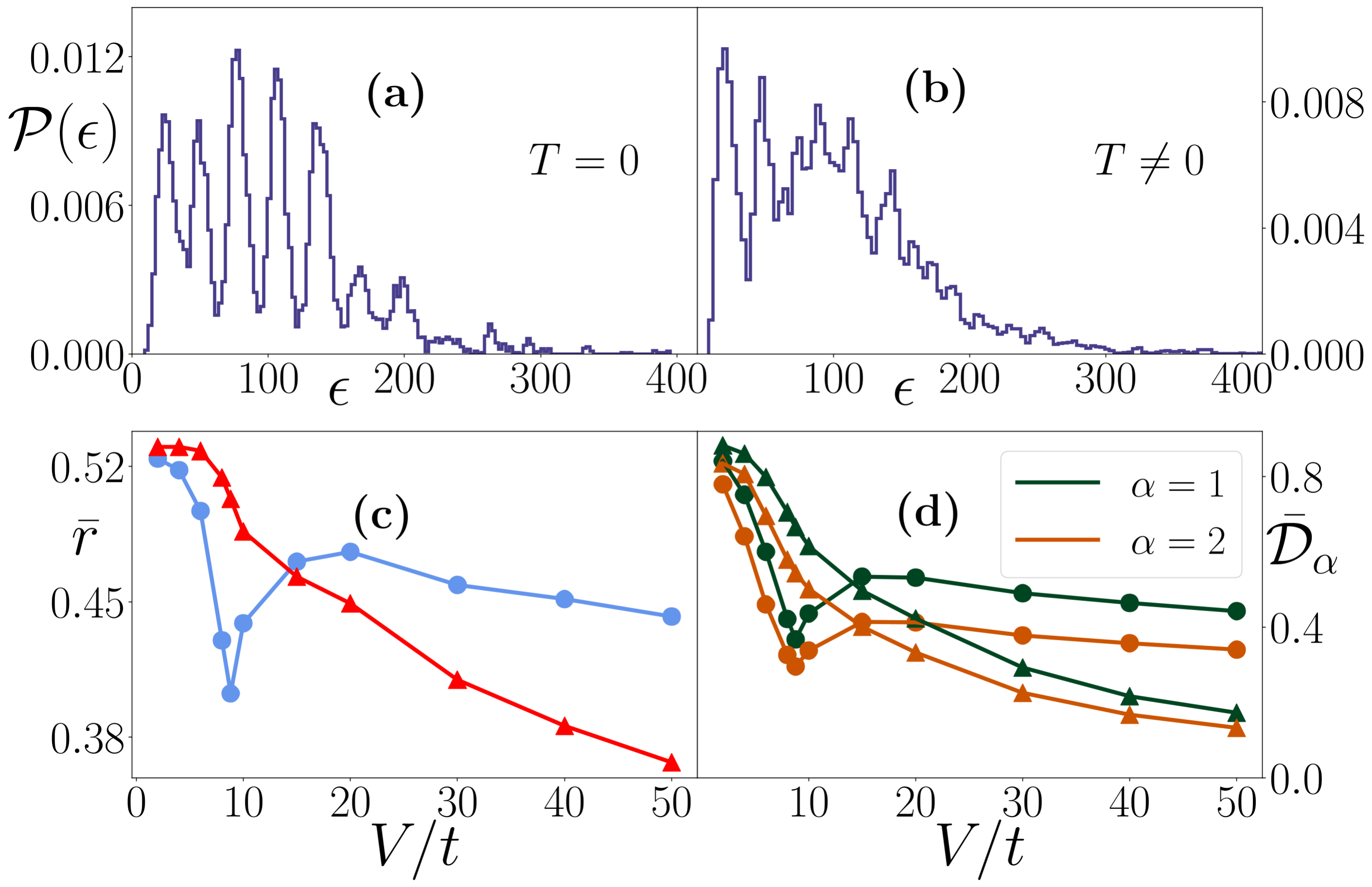}
\caption{Energy density for $T=0$~(a) and in the presence of IIT~(b) for $V/t=30$ and for $N=7$ bosons in $L=14$ sites. (c) Mean gap ratio, $\overline{r}$ and (d) mean fractal dimensions $\overline{\cal D}_\alpha$~of the eigenstates as a function of $V$ with~(circles) and without~(triangles) IIT. The figure has been adapted and reprinted
with permissions from \cite{Adith23} published in 2023 by the American Physical Society.}	
\label{fig:spect}
\end{figure}

The small system size makes it also possible to perform a spectral analysis, summarized in Fig.~\ref{fig:spect}. For sufficiently large $V$ the density of states, $\mathcal{P}(\epsilon)$, shows signatures of fragmentation in the form of pronounced peaks, responsible for the lack of full ergodicity. This structure is partially destroyed by IIT, explaining the enhanced localization. A common signature of global spectral properties is the mean gap ratio, $\overline r$, defined as an average of gap ratios, $r_n$:
\begin{equation}
r_n \equiv \frac{\min\lbrace\Delta_n,\Delta_{n+1}\rbrace}{\max\lbrace\Delta_n,\Delta_{n+1}\rbrace},
\end{equation} 
where $\Delta_n = \epsilon_{n+1}-\epsilon_{n}$ are the spacings between subsequent eigenenergies. For truly ergodic system, following random matrix theory predictions, $\overline r\eqsim 0.53$ while for the orderly, integrable case, $\overline r\simeq  0.389$~\cite{Atas13}. The mean gap ratio, for a model without IIT, shows a monotonic decrease with $V$ to the integrable value. In the presence of IIT we observe a sharp minimum around $T=-t$, the point where the negative interference of tunnelings occurs - c.f. Fig.~\ref{fig:spect}(c) - as well as mixed statistics even for the largest values of $V/t$ considered. These results are additionally confirmed by the fractal dimensions of the eigenstates, ${\cal D_\alpha}$, defined as ${\cal D}_\alpha = S_\alpha/{\cal N}$, where ${\cal N}$ is the Hilbert space dimension and

\begin{align}
S_\alpha = \frac{1}{1-\alpha}{\rm ln}\left (\sum_{i=1}^{\cal N} |\langle i |\Psi \rangle|^{2\alpha}\right),
\label{entropy}
\end{align}

are the participation entropies - compare  Fig.~\ref{fig:spect}(d). Since ${\cal D_1} \ne  {\cal D_2}$, the eigenstates seem to be multifractal, their higher values for the model which includes IIT for sufficiently large $V/t$ supports delocalization in that case.   

While the results reviewed for the full soft-core bosons case are obtained with exact diagonalization for small system sizes, they are expected to hold for larger systems as well as in 2D. In particular, the effect of negative interference of kinetic tunnelings with IIT~\cite{Kraus20,Adith23} should be amenable to experiments.

\section{Experimental realizations}
\label{sec:exp}

\subsection{Periodically modulated contact-interacting systems}

Contact-interacting systems usually are sufficiently described via the standard Bose-Hubbard Hamiltonian, as all additional terms typically are orders of magnitude smaller than the main terms on-site interactions $U$ and single-particle tunneling $t$. Nonetheless, under certain conditions this does not hold anymore, and additional extensions have to be taken into account to accurately describe the physics at play. Often, dynamical modifications of the lattice structure or of the interactions are key to induce those additional terms. 

\begin{figure}[t]
\includegraphics[width=1.0\columnwidth]{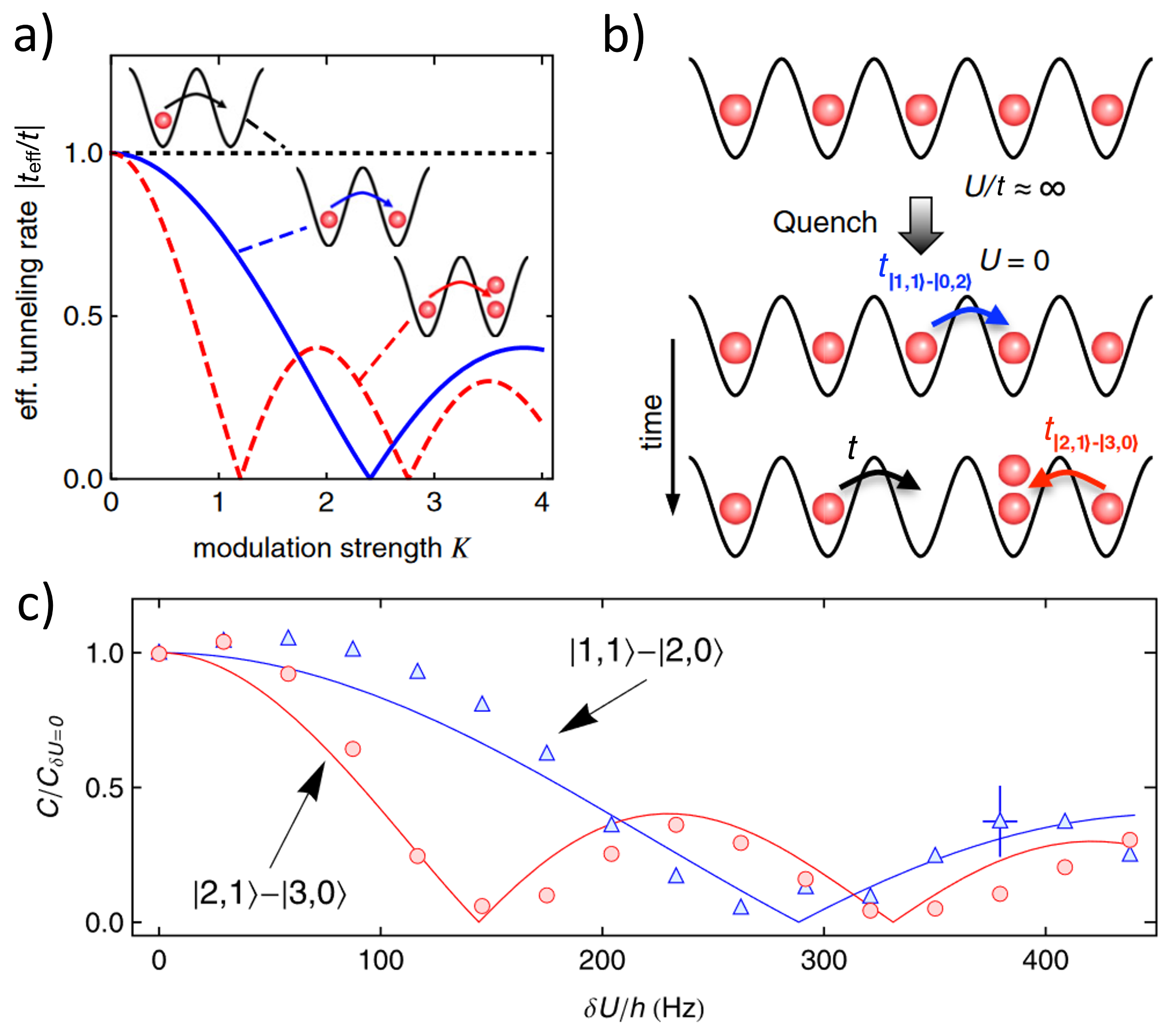}
\caption{Floquet engineering. a) Effective tunneling rate as a function of the modulation strength for two different occupation scenarios. b) Measurement procedure to detect the modified tunneling rates. c) Measured tunneling strengths normalized to the bare single-particle tunneling rate for the two processes indicated in a).
The figure has been adapted and reprinted with permissions from~\cite{Meinert2016} published in 2016 by the American Physical Society.}
\label{fig:Floquet}
\end{figure}

An example of such a situation comes from Floquet engineering described as a second example in Section \ref{induced} and realized via Eq.~\eqref{driven}. One successful experimental realization of this concept used cesium atoms in an optical lattice~\cite{Meinert2016}, which enables modulation of $U$ via its magnetic scattering length tunability~\cite{Mark2012}. To measure the effective tunneling rate of the system, they investigated the response of a singly-occupied Mott insulator to a quench in the on-site interaction, see Fig.~\ref{fig:Floquet}. More specifically, it was first quenched to zero $U=0$ to induce tunneling dynamics, and then -- after some evolution time -- quenched to a deep lattice $J\approx0$ to freeze the system. The observed reduction of singly occupied sites during this evolution time due to an increase of doubly- and triple-occupied sites matched the expected single-particle tunneling rate. Now they modulate the on-site interaction with varying $\delta U=U_1/\omega$ and observe a clear Bessel-type dependence of the tunneling dynamics on the modulation strength with pronounced minima indicating coherent destruction of tunneling. By using an additional gradient similar to earlier work~\cite{Meinert2013} they were able to independently measure the dependence on the occupation difference, see Fig.~\ref{fig:Floquet}.

Combining periodic modulation of the interaction strength with the periodic modulation of the bare tunneling rates via lattice shaking has also been experimentally demonstrated recently in~\cite{Clark2018ood}, realizing a density-dependent gauge field. Here, they again use a BEC of cesium atoms, now loaded into a single 2D square lattice plane with spacing $d=532\,$nm. The phase and therefore the lattice site positions of each lattice direction can be modulated independently, enabling full control over the relative phase $\theta_s$. The shaking frequency has been chosen to be slightly higher than the excitation gap. The resulting effective single-particle dispersion relation - for modulation amplitudes larger than a critical strength - develops four minima at finite momentum and induces a phase transition in which the condensate segregates into domains, each containing atoms occupying one of the minima. 

This picture stays only valid as long as interactions are not included. As the resulting micro-motion within a single lattice site modulates the onsite density, the onsite interaction strength gets modulated as well. When averaging this effect over a modulation period it can be shown that in general the modulation does not cancel out and shifts compared to the bare interaction strength. This means that the interaction energy develops a dependence on the direction of modulation, i.e. an interaction-momentum coupling, breaking the four-fold symmetry of the effective single-particle dispersion relation. Only for circular modulation $\theta_s=\pi/2$, this effect again vanishes. In this case, a synchronized modulation of the bare interaction strength with relative phase $\theta_g$ allows to reestablish the density-dependent gauge field.

A different scenario is realized when interactions are becoming comparable to the bandgap. Such a system has been realized again by using cesium atoms in an optical lattice, but pushing the interactions to large positive or negative values~\cite{Mark2020}. In this case, also higher-order terms, normally neglected, and dissipative on-site terms have been taken into account. The full Hamiltonian of the system then reads

\begin{align}
\hat{H} = &-t\sum_{i}\left(\hat{b}^\dagger_i\hat{b}_{i+1}+{\rm H.c.}\right)+\frac{U}{2}\sum_{i}\hat{n}_i(\hat{n}_i-1)\nonumber\\
&+\tilde{U}\sum_{i}\left(\hat{b}^\dagger_i\hat{b}^\dagger_i\hat{b}_i\hat{b}_{i+1}+\hat{b}^\dagger_i\hat{b}^\dagger_{i+1}\hat{b}_{i+1}\hat{b}_{i+1}+{\rm H.c.}\right)\nonumber\\
&-i\frac{\gamma_3}{12}\sum_{i}\hat{b}^{\dagger 3}_i\hat{b}^3_i
\label{occdeptun2}
\end{align}

Here, $\gamma_3$ denotes the corresponding three-body loss coefficient (in principle scaling with the fourth power of the scattering length), and $\tilde{U}$ the nearest-neighbor two-body interaction arising purely from contact interactions. For the contact interaction terms, the analytic corrections~\cite{Busch1998} due to the renormalization have been included as well. In the experiment, they are probing atom loss from a doubly occupied Mott insulator after a quench in the interaction parameter. They observe a peculiar scaling of the loss as a function of the scattering length with a strong asymmetry in positive vs. negative scattering lengths. While the overall behavior can be captured by the theory, a more quantitative description fails. The paper speculates that the effects of mixing ground and excited on-site three-body states and a further renormalization of the tunneling rates would be necessary to reproduce the experimental dynamics.

\subsection{Dipolar long-range systems}

This section deals with the experimental realization of spinless dipolar particles in optical lattice configurations, a recent review of experiments with magnetic quantum gases including fermionic gases and spinful Hamiltonians is given in~\cite{Chomaz2023}.

\begin{figure}[t]
\includegraphics[width=1.0\columnwidth]{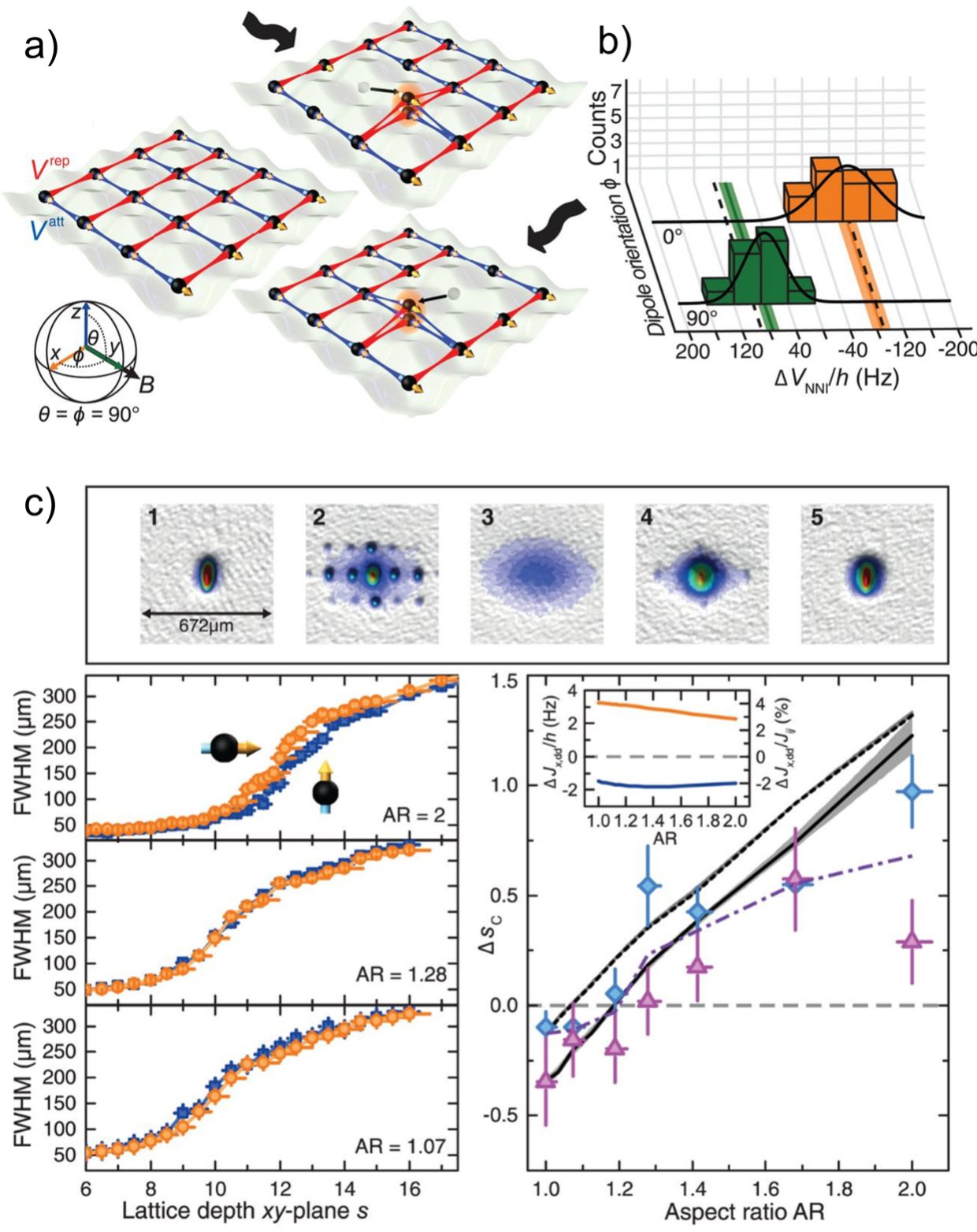}
\caption{Extended Bose–Hubbard model with dipolar bosons. a) Detection scheme for the nearest neighbor interaction using directional particle-hole excitations. b) Experimental measurement of the nearest neighbor interaction. c) Experimental evidence of an anisotropic shift of the superfluid to Mott insulator transition caused by anisotropic density-induced tunneling. The figure has been adapted and reprinted with permissions from~\cite{Baier2016} published in 2016 by the American Association for the Advancement of Science.}
\label{fig:eBHM3DLattice}
\end{figure}

The experimental realization of Bose-Hubbard models including dipolar extensions was a long-standing challenge. The reasons behind this were manifold: First, the most promising candidate to realize such models -- heteronuclear ground-state molecules yielding a large electric dipole moment -- turned out to be much more difficult to prepare with a reasonable phase space density or lattice occupation, as even for chemically stable combinations fast collisional losses occurred. This limited the realized Hamiltonians to spinful systems in a frozen regime to avoid collisions and subsequent losses. Only very recent  advances in experimental techniques have shown a workaround which allows efficient shielding of losses~\cite{Schindewolf2022}, which will open again the doors for future use of heteronuclear molecules for Bose-Hubbard physics. Second, magnetic atoms -- while brought to degeneracy without big hurdles on the way already more than 10 years ago~\cite{Griesmaier2005,Lu2011,Aikawa2012} -- feature only a comparatively weak dipolar strength and it was a priori not clear if the expected effects will be strong enough to be detected at realistic experimental parameters~\cite{Zhang2015}.

\begin{figure}[t]
\includegraphics[width=1.0\columnwidth]{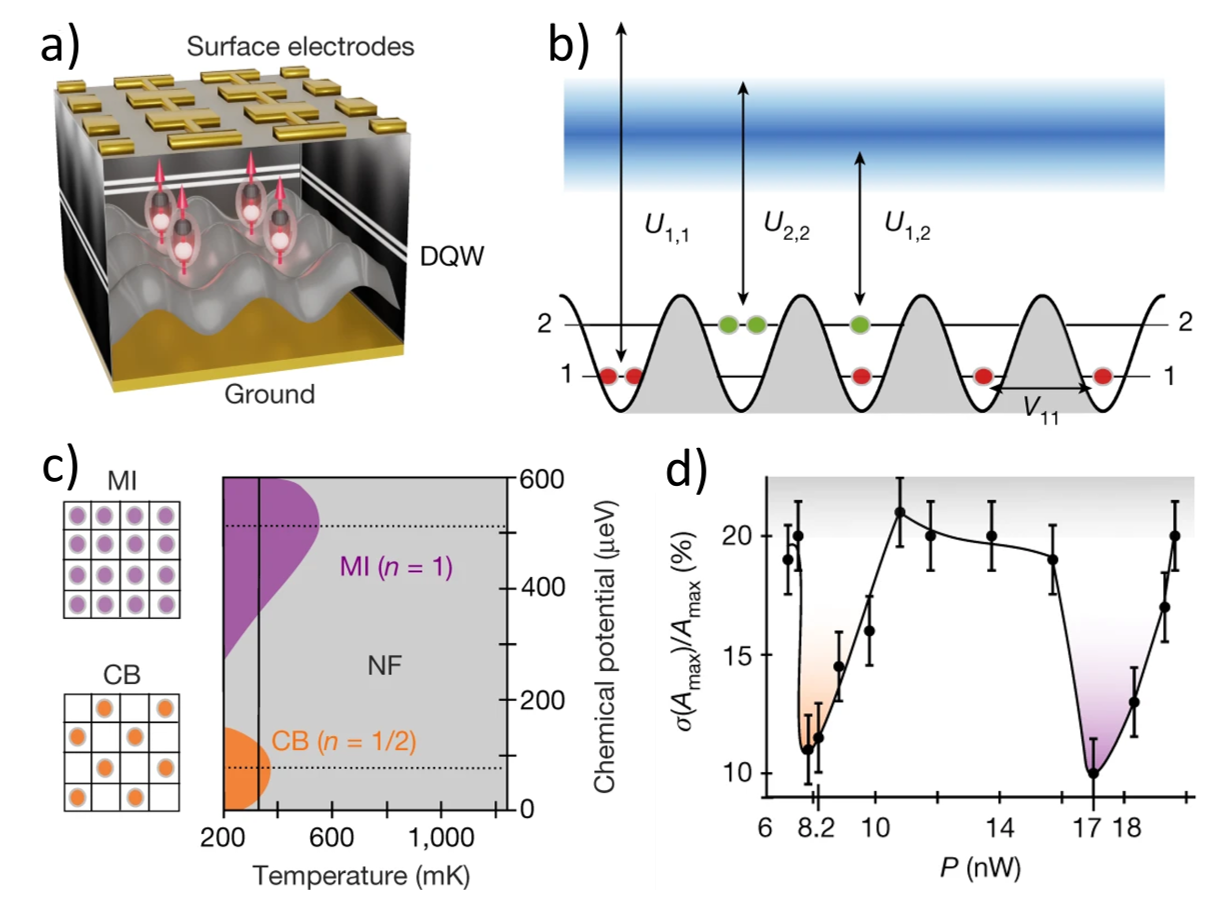}
\caption{Extended Bose–Hubbard model with dipolar excitons. a) Illustration of the physical system. b) Relevant energy terms of the realized Hubbard system. c) Expected phase diagram with the checkerboard phase at half filling. d) Measured compressibility of the system, indicating incompressible phases at half and unity filling. The figure has been adapted and reprinted with permission from~\cite{Lagoin2022} published in 2022 by the Nature Publishing Group.}
\label{fig:eBHMExcitons}
\end{figure}

Nonetheless, in 2016, the first experiment reported on the observation of dipolar long-range effects using spin-polarized bosonic $^{168}$Er atoms in a three-dimensional optical lattice~\cite{Baier2016}. A short lattice spacing along two directions, forming a tetragonal unit cell with spacings $(266,266,512)\,$nm, and the large magnetic moment of $7\,\mu_\text{B}$ resulted in a comparatively large nearest-neighbor interaction of $V/h\approx30\,$Hz in side-by-side configuration and $V/h\approx-60\,$Hz in head-to-tail orientation within planes. After adiabatic loading of a BEC of $^{168}$Er into the lattice and forming a Mott-insulating state, lattice modulation spectroscopy~\cite{Kollath2006} was used to map out the excitation spectrum as a function of lattice depths and dipole orientation relative to the (anisotropic) on-site wavefunction. This revealed the contribution of DDI to the on-site interaction energy which vanishes for a spherical symmetric situation. Doublon-hole excitations in the Mott-insulator can also change the number of attractive and repulsive nearest-neighbor bonds when the dipoles are oriented along one lattice direction, see Fig.~\ref{fig:eBHM3DLattice}. By using a differential measurement method, Baier et al. managed to experimentally determine the nearest-neighbor interaction energy difference between head-to-tail and side-by-side configurations to $\Delta V/h=80.5(17)\,$Hz. Finally, they characterized the angle-dependence of the quantum phase transition between superfluid and Mott-insulator. Here, not only the on-site contribution of DDI plays a significant role, but they found better agreement with theory when including density-induced tunneling being modified by DDI.

The extended Bose-Hubbard model was recently also realized using dipolar excitons~\cite{Lagoin2022}, a quasiparticle formed by an electron-hole pair in a semiconductor~\cite{Combescot2017}. The lattice is created by electric fields from an array of electrodes, forming a sinusoidal two-dimensional square lattice with $250\,$nm period. Excitons are then optically injected with laser pulses, whose power controls the mean density per lattice site. After they thermalize within a few nanoseconds they occupy essentially two Wannier states with large on-site interactions compared to tunneling ($U\gg t$) and comparatively strong nearest-neighbor interactions ($V/t\approx 20$). From photoluminescence spectra, Lagoin et al. deduced the compressibility and found an incompressible state at unity filling, corresponding to a Mott-insulator, see Fig.~\ref{fig:eBHMExcitons}. Here they additionally observe a shift in energy compared to very low fillings, in good agreement with the energy given by four nearest neighbor bonds $4V$. They also found such an incompressible state at half filling, indicating the preparation of a checkerboard phase.


The first single-site resolved observation of dipolar quantum solids has been realized very recently using a quantum gas microscope for erbium~\cite{su2023dipolar}. Using a single plane with lattice spacings of $(266,266)\,$nm, again, NN couplings of $V/h\approx30\,$Hz ($V/h\approx-60\,$Hz) were reached. The target state is prepared by adiabatically ramping the lattice depth up. The final imaging is done by freezing out the motion of atoms, expanding the density pattern with an accordion lattice and performing single-site resolved ultra-fast imaging~\cite{su2024fast}. Here, two counter-propagating imaging beams resonant with the main cooling transition of erbium at $401\,$nm - pulsed in alternating order - illuminate the atoms in order to scatter the maximum amount of photons within a few $\mu$s. This results in stochastic momentum kicks during this time and a diffusive broadening of the atom position. The favorable combination of a broad transition (= fast scattering rate), small wavelength (= high resolution) and large mass (= small momentum kicks) in lanthanides as erbium makes them especially applicable to such an imaging scheme. The final detection fidelity reaches above $99\%$ and additionally allows a parity-projection-free imaging~\cite{su2024fast}.

For half filling, the resulting ground states~\cite{Goral2002,Menotti2007,CapogrossoSansone2010,Zhang2015,Wu2020} are sensitively depending on the dipole orientation, with $\theta$ the polar angle and $\phi$ the azimuthal angle. For dipoles oriented perpendicular to the plane ($\theta=0^\circ$) they observe a checkerboard solid, while dipoles oriented along one lattice direction ($\theta=90^\circ,\phi=0^\circ$) results in a stripe phase. For diagonal orientation ($\theta=50^\circ,\phi=45^\circ$) the system exhibits diagonal ordering. This diagonal ordered pattern changes with the polar angle until at ($\theta=90^\circ,\phi=45^\circ$) phase separation happens with a central (elliptic) area with unity filling. The experiment also probed out-of-equilibrium dynamics by ramping from the superfluid into the phase-separated state ($\theta=90^\circ,\phi=45^\circ$) with varying ramp speed, showing that at fast ramps meta-stable diagonal stripes form, while adiabatic ramps lead to the phase-separated state.

\subsection{Cavity-enhanced systems}

As we already discussed in section~\ref{sec:cav}, long-range interactions and therefore extended Hubbard models can be also introduced by harnessing the back-action of a cavity onto the atoms inside of it. Specifically, light scattering from atoms into the cavity mode and back introduce an effective light-mediated up-to-infinite-range interaction that can be controlled independently to the other system parameters.

\begin{figure}[t]
\includegraphics[width=1.0\columnwidth]{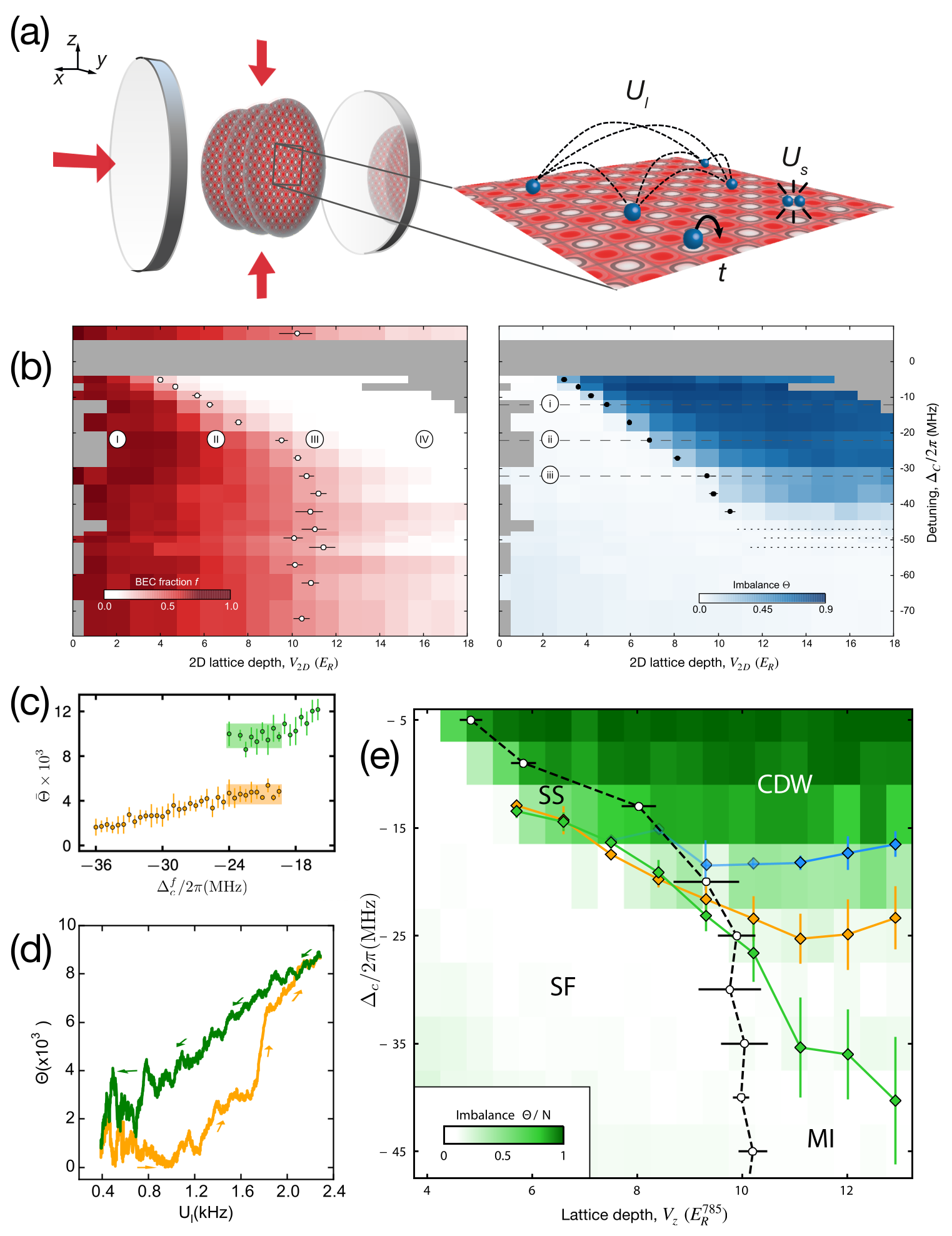}
\caption{Extended Bose–Hubbard model with cavity. (a)~Illustration of the experimental setup indicating the 2D lattice structure relative to the cavity. (b) Experimentally determined phase transitions. (c) Hysteresis in the imbalance when quenching close to the phase transition point. (d) Hysteresis in the MI-CDW phase transition observed for the different ramp directions. (e) Full phase diagram including the information where the hysteresis area opens (green and orange diamonds) and where imbalance jumps occur (blue diamonds)~\cite{Hruby2018}.
The panels (a) and (b) have been adapted and reprinted with Authors' permission from~\cite{Landig2016} published in 2016 by the Nature Publishing Group, and the panels (c)-(e) have been adapted and reprinted with Authors' permission from~\cite{Hruby2018} published in 2018 by the 
National Academy of Sciences.
}
\label{fig:eBHMCavity1}
\end{figure}

Such a system has been realized in Ref.~\cite{Landig2016}, see also Figure~\ref{fig:eBHMCavity1}. Here, they realized a lattice model with on-site and infinite range interactions mediated by the cavity photons. For this, they prepared a BEC of $^{87}$Rb atoms within an optical cavity with a Finesse of more than $\mathcal{F}>10^5$~\cite{Baumann2010}. Then they split up their system into separated 2D layers by an optical lattice formed by a back-reflected beam at $\lambda=670\,$nm propagating perpendicular to the cavity. Finally they create a 2D square lattice within each layer formed by one back-reflected beam at $\lambda=785\,$nm, again propagating perpendicular to the cavity, and light at $\lambda=785\,$nm co-linear with the cavity mode. The perpendicular $785\,$nm lattice is also responsible for inducing long-range interactions via off-resonant scattering of its photons into the cavity mode. The realized Hamiltonian then reads as:
\begin{align}
\hat{H} = &-t\sum_{\langle e,o\rangle}\left(\hat{b}^\dagger_e\hat{b}_{o}+{\rm H.c.}\right)+\frac{U_s}{2}\sum_{i}\hat{n}_i(\hat{n}_i-1)\nonumber\\
&-\frac{U_l}{L}\left(\sum_{e}\hat{n}_e-\sum_{o}\hat{n}_o \right)^2-\sum_{i}\mu_i\hat{n}_i
\label{occdeptun3}
\end{align}

The infinite-range interaction, given by $U_l$, acts between even ($e$) and odd ($o$) sites and -- for positive $U_l$ -- favors a particle imbalance between the two checkerboard sublattices located on even and odd sites respectively. It can be controlled independently by the detuning of the perpendicular $\lambda=785\,$nm lattice light with respect to the cavity resonance frequency. To characterize the system, they measure the presence of global phase coherence by probing the interference pattern of the cloud after free expansion. This information indicates the transition between a superfluid and an insulating state. Additionally they record the amplitude of the scattered light leaking out of the cavity. This amplitude is directly proportional to the population imbalance between even and odd sites, indicating a transition from a regular Mott insulator to a charge density wave, or from a superfluid to a supersolid.

Figure~\ref{fig:eBHMCavity1} shows the experimentally obtained phase transitions using the two probes, which can be used to construct the full phase diagram as shown in Fig.~\ref{fig:cavity_setup_exp}. For large detunings (small $U_1$) they observe the usual superfluid-to-Mott insulator phase transition, while for small detunings (large $U_1$) they detect that the Mott insulator is replaced by a charge density wave state, and that a supersolid state -- sharing both superfluid and charge density wave properties -- lies in between the isolating state and the superfluid state. They also observed hysteresis appearing when performing ramps between the Mott insulator and the charge density wave state, pointing towards a first-order phase transition between them.

This also shows that the system can be used to explore out-of-equilibrium dynamics within this Hamiltonian. In a subsequent work~\cite{Hruby2018} the group investigated the initially observed hysteresis and additional quench dynamics in more detail. Especially the availability of real-time information on the system obtained from the scattered light signal allows an in-depth analysis. Using quench experiments they observe a metastable region where the measured imbalance depends on the starting state of the quench. They characterize this region by recording the time-resolved imbalance when ramping over the phase transition and determining the turning points of the imbalance against the ramp value. They also observe a peculiar jump in the time-resolved imbalance data when ramping from a Mott insulator into the charge density wave state. This behavior could be explained by atoms within the central Mott insulator tunneling collectively to even (odd) sites and thereby building up the charge density wave. This collective event gets triggered by individual tunneling events happening first in the outer layers due to the experimental harmonic trapping, increasing the mobility at the outer areas of the system. 

\section{Conclusions}

In this report on progress we collected some of the most recent and exciting results on the investigation of non-standard Bose-Hubbard models in the context of atomic quantum simulators. Specifically, we first provided a general derivation of these iconic Hamiltonians and subsequently discussed how different interacting terms and/or geometrical configurations can give rise to intriguing states of matter and quantum mechanical effects. Notably, we tackled these subjects both from a theoretical and an experimental perspective.

Specifically to the theory side, we covered topics ranging from beyond Landau's criticality and the role of interaction induced tunnelings  to cavity mediated interactions and dipolar systems in out-of-equilibrium configurations. The experimental sections have  described recent setups made of ultracold atoms with special attention on Floquet engineerings, phases of matter induced by strong dipolar repulsion and atom-cavity setups.

This impressive amount of recent results which were not present in the last review on this fascinating subject \cite{Dutta15}, demonstrates again the central importance that non-standard Hubbard models had, have and will have in order to understand fundamental laws of nature appearing in condensed matter and, as recently realized, high energy physics \cite{Yang2020} and quantum chemistry \cite{Luengo2019}. We thus believe that this review poses the ground towards many more scientific achievements that the future will bring.


\section{Acknowledgments}
We are deeply grateful to all the colleagues which have collaborated with us on the subject treated in this review.
We are most grateful to Tobias Donner for providing us with their original drawings for Fig.~\ref{fig:cavity_setup_exp}(b) and Fig.~\ref{fig:eBHMCavity1}. We thank Francois Dubin, Francesca Ferlaino, Benoit Gremaud, and Giovanna Morigi for permissions to reproduce figures from the articles they coauthored. 
L.B. acknowledges financial support within the DiQut Grant No.
2022523NA7 funded by European Union – Next
Generation EU, PRIN 2022 program (D.D. 104
- 02/02/2022 Ministero dell’Università e della
Ricerca).
M.L. acknowledges support from: ERC AdG NOQIA; Ministerio de Ciencia y Innovation Agencia Estatal de Investigaciones (PGC2018-097027-B-I00/10.13039/501100011033, CEX2019-000910-S/10.13039/501100011033, Plan National FIDEUA PID2019-106901GB-I00, FPI, QUANTERA MAQS PCI2019-111828-2, QUANTERA DYNAMITE PCI2022-132919, Proyectos de I+D+I “Retos Colaboraci\'on” QUSPIN RTC2019-007196-7); MICIIN with funding from European Union NextGenerationEU(PRTR-C17.I1) and by Generalitat de Catalunya; Fundació Cellex; Fundaci\'o Mir-Puig; Generalitat de Catalunya (European Social Fund FEDER and CERCA program, AGAUR Grant No. 2021 SGR 01452, QuantumCAT \ U16-011424, co-funded by ERDF Operational Program of Catalonia 2014-2020); Barcelona Supercomputing Center MareNostrum (FI-2022-1-0042); EU Horizon 2020 FET-OPEN OPTOlogic (Grant No 899794); EU Horizon Europe Program (Grant Agreement 101080086 — NeQST), ICFO Internal “QuantumGaudi” project; European Union’s Horizon 2020 research and innovation program under the Marie-Skłodowska-Curie grant agreement No 101029393 (STREDCH) and No 847648 (“La Caixa” Junior Leaders fellowships ID100010434: LCF/BQ/PI19/11690013, LCF/BQ/PI20/11760031, LCF/BQ/PR20/11770012, LCF/BQ/PR21/11840013).  
L. B. acknowledges financial support within the DiQut Grant No. 2022523NA7 funded by European Union – Next Generation EU, PRIN 2022 program (D.D. 104 - 02/02/2022 Ministero dell’Universit`a e della Ricerca).
M.J.M. acknowledges financial support from a NextGeneration EU Grant AQuSIM through the Austrian Research Promotion Agency (FFG) (No.\,FO999896041), and the Austrian Science Fund (FWF) Cluster of Excellence QuantA (\href{https://doi.org/10.55776/COE1}{10.55776/COE1}).
This research was also funded by National Science Centre (Poland) under the OPUS call within the WEAVE programme 2021/43/I/ST3/01142 (J.Z.). A partial support by the Strategic Programme Excellence Initiative at Jagiellonian University as well as that within the QuantEra II Programme that has received funding from the European Union's Horizon 2020 research and innovation programme under Grant Agreement No 101017733 DYNAMITE (M.L. and J.Z.).

Views and opinions expressed in this work are, however, those of the authors only and do not necessarily reflect those of the European Union, European Climate, Infrastructure and Environment Executive Agency (CINEA), nor any other granting authority. Neither the European Union nor any granting authority can be held responsible for them.

%

\end{document}